\documentclass[12pt]{article}
\pdfoutput=1
\usepackage[square,comma,numbers,sort&compress]{natbib}
\usepackage{graphicx,epstopdf,amssymb,amsfonts,amsmath,amsthm,array,
mathrsfs,amscd}
\usepackage[vcentermath]{youngtab}
\usepackage{float}
\usepackage{hyperref}
\newcommand{\pbs}[1]{\let\temp=\\#1\let\\=\temp}
\numberwithin{equation}{section}
\def\be{\begin{equation}}\def\ee{\end{equation}}
\def\cvp{\raise 2pt\hbox{,}} 
 \def\tr{\mathop{\rm tr}\nolimits}
\def\im{\mathop{\rm Im}\nolimits}
\def\re{\mathop{\rm Re}\nolimits}  

 \def\d{{\rm d}}\def\nn{{\cal
N}} 
 \def\uN{\text{U}(N)}\def\uK{\text{U}(K)}

\def\la{\lambda}\def\La{\Lambda}
\def\uNK{\text{U}(N+K)}
\def\plb#1#2#3{{\it Phys.\ Lett.\ }{\bf B #1} (#2) #3}
\def\npb#1#2#3{{\it Nucl.\ Phys.\ }{\bf B #1} (#2) #3}

\def\prl#1#2#3{{\it Phys.\ Rev.\ Lett.\ }{\bf #1} (#2) #3}
\def\jhep#1#2#3{{\it J. High Energy Phys.\ }{\bf #1} (#2) #3}
\def\prd#1#2#3{{\it Phys.\ Rev.\ }{\bf D #1} (#2) #3}

\def\atmp#1#2#3{{\it Adv.\ Theor.\ Math.\ Phys.\ }{\bf #1} (#2) #3}
\def\cmp#1#2#3{{\it Comm.\ Math.\ Phys.\ }{\bf #1} (#2) #3}

\def\imath#1#2#3{{\it Invent math }{\bf #1} (#2) #3}

\begin{document}
%
%
{\pagestyle{empty}
\parskip 0in
\

\vfill
\begin{center}
{\LARGE D-Brane Probes in the Matrix Model}



\vspace{0.4in}

Frank F{\scshape errari}
\\
\medskip
{\it Service de Physique Th\'eorique et Math\'ematique\\
Universit\'e Libre de Bruxelles and International Solvay Institutes\\
Campus de la Plaine, CP 231, B-1050 Bruxelles, Belgique}
\smallskip
{\tt frank.ferrari@ulb.ac.be}
\end{center}
\vfill\noindent

Recently, a new approach to large $N$ gauge theories, based on a generalization of the concept of D-brane probes to any gauge field theory, was proposed. In the present note, we compute the probe action in the one matrix model with a quartic potential. This allows to illustrate several non-trivial aspects of the construction in an exactly solvable set-up. One of our main goal is to test the bare bubble approximation. The approximate free energy found in this approximation, which can be derived from a back-of-an-envelope calculation, matches the exact result for all values of the 't~Hooft coupling with a surprising accuracy. Another goal is to illustrate the remarkable properties of the equivariant partial gauge-fixing procedure, which is at the heart of the formalism. For this we use a general $\xi$-gauge to compute the brane action. The action depends on $\xi$ in a very non-trivial way, yet we show explicitly that its critical value does not and coincides with twice the free energy, as required by general consistency. This is made possible by a phenomenon of ghost condensation and the spontaneous breaking of the equivariant BRST symmetry.

\vfill

\medskip
%
\begin{flushleft}
\today
\end{flushleft}
\newpage\pagestyle{plain}
\baselineskip 16pt
\setcounter{footnote}{0}

}


%
\section{\label{s1Sec} Introduction}

\subsubsection*{The framework}

Recently, the author has proposed a new approach to large $N$ gauge theories \cite{ferfund}, synthesizing and extending ideas appearing in a series of previous works \cite{fer}. The method can a priori be applied to any gauge theory, in any number of dimensions, independently of supersymmetry or the existence of an explicit construction in standard string theory.

The basic idea is as follows. We focus, for concreteness, on a $\uN$ gauge theory in $p+1$ spacetime dimensions. This theory provides a notion of D$p$-branes: by definition, it governs the dynamics of a stack of $N$ D$p$-branes. Now consider a stack of $N+K$ branes, governed by the $\text{U}(N+K)$ gauge theory. The fundamental objects in the construction of \cite{ferfund} are actions $\mathscr A_{N,K}$ which describe the dynamics of $K$ branes in the stack of $N+K$, in the presence of the other $N$ branes. In the leading large $N$ and fixed $K$ limit, which we shall always consider in the following, the action takes the form $\mathscr A_{N,K}= NA_{K}$, where $A_{K}$ does not depend on $N$. It is then natural to call the $K$ branes we single out the ``probe'' branes and the $N$ other branes the ``background'' branes. The basic properties of the probe brane actions $A_{K}$ are the following.

\noindent (i) They are $\uK$ invariant and describe the dynamics of $K$ D-branes probing an emergent holographic bulk geometry which is dual to the original large $N$ gauge theory. In particular, they depend generically on a certain number of $\uK$ adjoint scalar fields corresponding to the coordinates of the holographic emergent space. This is true even when the gauge theory under consideration does not have any elementary scalar field, like in the case of the pure Yang-Mills model.

\noindent (ii) They are built using a partial gauge-fixing of the $\uNK$ gauge symmetry down to $\uN\times\uK$. This procedure involves an equivariant version of the usual BRST cohomology and produces quartic ghost terms at tree level \cite{ferequiv,Schaden}. The resulting actions $A_{K}=A_{K}[\psi]$ depend crucially on the choice of an equivariant gauge-fixing fermion $\psi$. This gauge-dependence of the construction is related to the fact that the local bulk physics is not observable.

\noindent (iii) The on-shell D-brane action $A_{K}[\psi]^{*}$, obtained by evaluating $A_{K}[\psi]$ on the solution of its classical equations of motion, is related to the planar free energy $N^{2}F$ of the original gauge theory by the identity
\be\label{AFrel} F = \frac{1}{2K}A_{K}[\psi]^{*}\, .\ee
In particular, the on-shell action $A_{K}[\psi]^{*}=A_{K}^{*}$ does not depend on $\psi$ (it is gauge invariant). If we denote by $A[\psi]=A_{1}[\psi]$ the action for a single brane, we must then also have
\be\label{AKArel} A_{K}^{*}=K A^{*}\, ,\ee
since the left-hand side of \eqref{AFrel} does not depend on $K$.

The planar free energy $F$ can be defined by including sources for arbitrary single-trace gauge invariant operators in the gauge theory action. It thus contains all the information about the large $N$ limit of the gauge theory. Eqs.\ \eqref{AFrel} and \eqref{AKArel} then show that the probe brane action $A[\psi]$ contains all the information about the original large $N$ gauge theory as well. The computation of $A[\psi]$ thus provides a new  approach to the large $N$ limit, coding the same information as, for example, the more traditional 1PI effective action. The crucial advantage of $A[\psi]$ is that it automatically makes the link with a holographic description of the gauge theory (this is point (i) above). Moreover, and this may be the most exciting point about the whole story, the action $A[\psi]$ is amenable to study within a new ``bare bubble'' non-perturbative approximation scheme described in \cite{ferfund}. This approximation is able, in principle, to capture both the weak-coupling (it is always one-loop exact) and the strong coupling regimes of the gauge theory. 

The aim of the present paper is to illustrate the above ideas on the simplest, albeit already quite non-trivial, $\uN$ gauge theory: the zero-dimensional Hermitian matrix model with action 
\be\label{SNdef} S_{N}(M) = \frac{N}{\la}\tr_{N} \Bigl(\frac{1}{2}M^{2} + \frac{1}{4}M^{4}\Bigr)\, .\ee
Of course, the exact solution of this model has been known for a long time \cite{MMsol}. The planar free energy $F$ as a function of the 't~Hooft coupling $\la$, or any planar correlator, can be computed exactly. These quantities display some very non-trivial behavior as a function of $\la$. For example, they admit a non-trivial strong-coupling expansion around $\la=\infty$, the strong coupling expansion parameter being $1/\sqrt{\la}$. The same qualitative behavior is found in much more complicated models. For example, in the four-dimensional $\nn=4$ gauge theory, the strong coupling expansion is governed by the $\alpha'$ corrections of string theory in $\text{AdS}_{5}\times\text{S}^{5}$ which, in terms of the 't~Hooft coupling, is also an expansion in $1/\sqrt{\la}$.

Having an exactly solvable toy model like \eqref{SNdef} at hand is very valuable. It allows us to check every detail of the D-brane probe formalism explicitly and to uncover several non-trivial features that we believe are generic and could play a r\^ole in much more realistic contexts. Moreover, by comparing with the exact solution, we are able to assess quantitatively the accuracy of the bare bubble approximation. It turns out that it yields a remarkably, even unreasonably, good approximation to the exact free energy. We also get a very nice illustration of the non-standard equivariant gauge-fixing procedure, which plays a central r\^ole in the construction of the D-brane action.

\subsubsection*{Holography}

The standard lore of holography in the matrix model is very simple. The holographic dimension is usually identified with the space of eigenvalues of the matrix $M$. Classically, all these eigenvalues sit at the classical minimum $M=0$ of the action \eqref{SNdef}, but when the planar diagrams are summed up they spread and fill an interval $[-a,a]$. This spreading is due to a repulsive force between eigenvalues that is generated by the Vandermonde determinant coming from the diagonalization of the matrix $M$ in the path integral measure. The region where the eigenvalues sit is the  ``deep IR'' region of the geometry. The size of this region is very small, or order $\la^{1/2}$, at weak 't~Hooft's coupling, but grows like $\la^{1/4}$ at strong coupling. 

This lore yields a very simple picture of the notion of D-brane probes. A probe is identified with an eigenvalue, moving in a quantum potential which is the sum of the classical potential \eqref{SNdef} with the repulsive term due to the presence of the $N$ other eigenvalues in the interval $[-a,a]$. Intuitively, this quantum potential should be naturally identified with the brane action that we have denoted by $A$ above.

We shall see that this picture does perfectly fit in our framework. It is, however, only a very special case of the general construction, corresponding to a very particular choice of equivariant gauge-fixing condition used to define the probe brane action. This special gauge choice is akin to a Landau-like gauge, which would be inconvenient (in particular non-renormalizable) in more realistic models. For more general gauge choices, the eigenvalue interpretation for the D-brane probe is lost. One then gets a very different holographic description, based on a very different-looking brane action $A$. The difference between all the possible descriptions is tiny in the UV region of the holographic dimension, far away from the background eigenvalues, but it becomes crucial in the deep bulk. As discussed extensively in \cite{ferfund}, this ambiguity in the bulk description is directly related to the problem of bulk spacetime locality in quantum gravity. We will compute $A$ as a function of the gauge-fixing and explicitly check that all the resulting ``complementary'' holographic descriptions are indeed strictly equivalent on-shell.

\subsubsection*{Equivariant gauge-fixing}

A crucial and subtle aspect of the construction of the D-brane actions is the procedure of partial gauge-fixing. For our purposes, it is enough to  consider the case of a single D-brane probe. The construction of the action then starts by considering the model \eqref{SNdef}, but now with $N+1$ colors instead of $N$. The $(N+1)\times (N+1)$ Hermitian matrix $M$ is decomposed according to the action of the $\uN\times\text{U}(1)$ subgroup of $\text{U}(N+1)$ as
\be\label{funddec} M = 
\begin{pmatrix} V^{a}_{\ b} & \bar w^{a}\\ w_{a} & v
\end{pmatrix}\, ,\ee
where the index $a$ runs from 1 to $N$. Intuitively, the D-brane probe action $A(v)$ should be obtained by integrating out $w$, $\bar w$ and $V$. However, when the $\text{U}(N+1)$ symmetry of the model is gauged, which we always assume here, we have to take into account the fact that the decomposition \eqref{funddec} is not invariant under $\text{U}(N+1)$ gauge transformations but only under the $\uN\times\text{U}(1)$ subgroup. In order to construct the D-brane probe action, we must thus supplement \eqref{funddec} with a partial gauge-fixing condition of the original gauge symmetry $\text{U}(N+1)$ down to $\uN\times\text{U}(1)$. 

The partial, equivariant, gauge-fixing procedure is non-standard and quite interesting in itself. It was recently developed from first principles in \cite{ferequiv} (see also \cite{Schaden} for earlier relevant references in the lattice literature). In particular, \emph{it is totally different from the familiar background field gauge-fixing procedure} which seems to have been used extensively in previous attempts to discuss D-brane actions.\footnote{The reasons why the background field gauge is \emph{not} the right procedure to define the D-brane action are explained in details in \cite{ferfund}.} One of the most salient feature of the formalism is to show that quartic ghost couplings must generically be included in the tree-level ghost Lagrangian. These quartic ghost terms have nothing to do with the familiar BRST-exact quartic ghost terms that are sometimes introduced in the standard BRST formalism for renormalization purposes but which, being BRST-exact, do not play any r\^ole in the computation of gauge-invariant observables. The quartic ghost terms of the partial gauge-fixing procedure are of an entirely different nature. They are present even in the matrix model, where obviously renormalization is not an issue, and are actually crucial to ensure the on-shell gauge invariance of the formalism.

The simplest partial gauge-fixing condition is to set the off-diagonal components of $M$ to zero,
\be\label{Landau} w = \bar w = 0\, .\ee
In this gauge, the lower-right corner $v$ of the matrix $M$ in \eqref{funddec} really is an eigenvalue. However, the Laudau-like gauge choice \eqref{Landau} is very special. For example, in multi-matrix models it cannot be applied to all the matrices. In higher dimensions, it would correspond to an awkward, non-renormalizable, gauge condition. It is thus both natural and required to understand more general partial gauge-fixing conditions. In the present paper, we shall use a general $\xi$-gauge, for which the gauge conditions \eqref{Landau} are not imposed sharply in the path integral but rather appear in a gauge-fixing term $w\bar w/\xi$ in the action. The resulting D-brane action $A[\xi](v)$ then depends explicitly on $\xi$ and the interpretation of $v$ as being an eigenvalue is lost.

Being able to deal with situations where the eigenvalue picture is lost is one of the virtues of the formalism. Even though the lower-right components $v$ in the decomposition \eqref{funddec} of the adjoint fields have been traditionally interpreted as eigenvalues, the precise meaning of this interpretation was always very unclear, in particular when several matrices are present, as in the $\nn=4$ model with its six scalar fields. We now see that this interpretation is usually misleading, even in the one matrix model! In particular, in four-dimensional gauge theories, natural renormalizable equivariant gauge choices will be Lorenz-like, as described in \cite{ferequiv}, with no eigenvalue interpretation for $v$.

One of the main goals of the present note is to illustrate explicitly how this gauge-dependence of the holographic picture works for the one matrix model. A crucial point is that all the actions $A[\xi]$, for all values of $\xi$, must be physically equivalent. This means that their on-shell critical values must all coincide with twice the planar free energy of the original matrix model, as dictated  by \eqref{AFrel} for $K=1$. In other words, if we solve the equation of motion
\be\label{eom} \frac{\d A[\xi]}{\d v} = 0\ee
and pick the solution $v=v^{*}[\xi]$ corresponding to the absolute mininum of $A[\xi]$, then
\be\label{Astar} A[\xi]\bigl(v^{*}[\xi]\bigr)=A[\xi]^{*}=A^{*}= 2F\, .\ee
It turns out that the quartic ghost terms are absolutely crucial for this on-shell gauge invariance to be valid. This property is actually far from being manifest on the exact solution that we shall derive below. Nevertheless, we can show explicitly that \eqref{Astar} is indeed satisfied, at the absolute minimum of $A[\xi](v)$ (but not for the other critical points). At this absolute minimum, the quartic ghost terms induce ghost condensation and the spontaneous breaking of the equivariant BRST symmetry. 

\subsubsection*{Plan of the paper}

In Section 2, we present the exact solution of the matrix model and briefly review the matrix model technology that we need. In Section 3, following \cite{ferfund}, we construct the probe brane action $A$ and discuss the associated equivariant BRST symmetry. In Section 4, we compute $A$ exactly. This allows to check explicitly several non-trivial properties of the formalism. We show that the equivariant BRST symmetry is spontaneously broken and that ghost condensation occurs. In Section 5, we compute $A$ in the bare bubble approximation. The approximate solution, which is obtained from a very simple calculation, reproduces very well the properties of the exact solution, even at the quantitative level. We conclude in Section 6.

\section{Brief review of matrix model technology\label{s2Sec}}

\subsubsection*{Basic definitions and relations}

The basic quantities that one wishes to compute in the model \eqref{SNdef} include the partition function
\be\label{ZNdef} Z_{N}(\la) =\la^{-N^{2}/2} \int\!\d M\, e^{-S_{N}(M)}\ee
or the associated free energy
\be\label{FNdef} \mathcal F_{N}(\la) = -\ln Z_{N}(\la)\ee
and the correlators
\be\label{CNdef} C_{N,k}(\la) = \frac{1}{N}\bigl\langle\tr_{N} M^{k}\bigr\rangle\, .\ee
We could also consider the generating function for all the correlators, which is the partition function \eqref{ZNdef} for an action $S_{N}$ which is the trace of an arbitrary polynomial in $M$. This can be done straightforwardly but, for simplicity, we shall concentrate on the quartic action \eqref{SNdef}.

We normalize the integration measure $\d M$ in such a way that
\be\label{normdM} \int\!\d M\, e^{-\frac{N}{2}\tr_{N}M^{2}}=1\, .\ee
The factor $\la^{-N^{2}/2}$ in the right-hand side of \eqref{ZNdef} then ensures that the partition function and the free energy have the standard perturbative expansions $Z_{N}(\la) = 1+O(\la)$, $\mathcal F_{N}(\la) = O(\la)$. In the leading large $N$ limit on which we focus, the free energy is given by $\mathcal F_{N}(\la)=N^{2}F(\la) + O(N^{0})$, where $F(\la)$, the planar free energy, is $N$-independent and given by an infinite perturbative series,
\be\label{Fpert} F(\la) = \sum_{p\geq 1}f_{p+1}\la^{p}\, .\ee
The coefficient $f_{p}$ is given by summing over the planar Feynman diagrams with $p$ loops. In the same leading large $N$ limit, the correlators $\eqref{CNdef}$ have been normalized in such a way that they have a smooth limit,
\be\label{Cinfi} C_{\infty,k}(\la) = C_{k}(\la)\, .\ee
The correlators can be conveniently encoded in the resolvent
\be\label{gdef} g(z;\la) = \lim_{N\rightarrow\infty}\frac{\la}{N}\Bigl\langle\tr_{N}\frac{1}{z-M}\Bigr\rangle\, ,\ee
whose large $z$ expansion reads
\be\label{gasymp} g(z;\la)= \la\sum_{k\geq 0}\frac{C_{k}(\la)}{z^{k+1}}\,\cdotp\ee
The free energy can actually be straightforwardly derived from the correlators, since by taking the derivative of \eqref{ZNdef} with respect to $\la$ we get
\be\label{FversusCorr} F'(\la) = \frac{1}{2\la} - \frac{C_{2}(\la)}{2\la^{2}} - \frac{C_{4}(\la)}{4\la^{2}}\, \cdotp\ee
By rescaling $M\mapsto\sqrt{\la}M$ in \eqref{ZNdef}, we can show similarly that
\be\label{FversusCorr4} F'(\la) = \frac{C_{4}(\la)}{4\la^{2}}\,\cdotp\ee
Inverting \eqref{FversusCorr} and \eqref{FversusCorr4} we also get the interesting relations
\be\label{FtoCorr} C_{2}(\la) = \la - 4\la^{2}F'(\la)\, ,\quad C_{4}(\la) = 4\la^{2}F'(\la)\, .\ee

\subsubsection*{The exact solution}

The solution of the model can be naturally expressed in terms of a density of eigenvalues $\rho(x;\la)$, which plays the r\^ole of a large $N$ master field for the matrix $M$. This density has support on an interval $[-a(\la),a(\la)]$. Any planar correlator can be expressed straightforwardly in terms of $\rho$,
\begin{align}\label{Ckrho}  C_{k} & = \int_{-a}^{a}x^{k}\rho(x)\,\d x\, ,\\\label{grho}  g(z) & = \la\int_{-a}^{a}\frac{\rho(x)}{z-x}\,\d x\, .
\end{align}
(We do not always indicate explicitly the dependence in $\la$). Of course, due to the $\mathbb Z_{2}$ symmetry $M\mapsto -M$ of the action \eqref{SNdef}, the density $\rho$ is an even function of $x$ and the correlators of the form $C_{2k+1}$ all vanish. Eq.\ \eqref{grho} shows that $g(z)$, when analytically continued on the complex $z$-plane, must have a branch cut on the interval $[-a,a]$. The discontinuity across the branch cut is related to the density of eigenvalue,
\be\label{gdisc} \rho(x)=\frac{i}{2\pi\la}\Bigl( g(x+i\epsilon) - g (x-i\epsilon)\Bigr)\, ,\quad x\in [-a,a]\, .\ee

The density must satisfy the so-called saddle point equation
\be\label{saddlerho} 2\la\, \text{P}\!\!\int_{-a}^{a}\frac{\rho(x')}{x-x'}\,\d x' = x^{3}+x\, ,\quad \text{for}\ x\in [-a,a]\, ,\ee
where the symbol $\text{P}$ means that we take the principal value of the integral. When expressed in terms of the resolvent, this equation is equivalent to the condition
\be\label{saddleg} g(x+i\epsilon) + g(x-i\epsilon) = x^{3}+x\, ,\quad \text{for}\ x\in [-a,a]\, .\ee
This shows that $g(z)$ must be a two-sheeted analytic function and in particular must satisfy a degree two algebraic equation. Together with the one-cut structure of $g$ on the $z$-plane and its asymptotics
\be\label{gasym} g(z)\underset{z\rightarrow\infty}{\sim}\frac{\la}{z}\, \cvp\ee
which is valid on the ``first'' sheet on which the expansion \eqref{gasymp} is valid, this allows to fix the solution uniquely. 

One finds that the size $a$ of the branch cut satisfies the algebraic equation
\be\label{aalgeq} a^{2}(3a^{2}+4)=16\la\, .\ee
Since $a$ must go to zero when $\la$ goes to zero, this yields
\be\label{asol} a^{2} = \frac{2}{3}\Bigl(\sqrt{1+12\la} -1\Bigr)\, .\ee
The resolvent itself is given by
\be\label{gsol} g(z) = \frac{z^{3}+z}{2}-\frac{1}{2}\Bigl(z^{2}+1+\frac{a^{2}}{2}\Bigr)\sqrt{z^{2}-a^{2}}\, ,\ee
where the square root is defined in such a way that the branch cut is on the interval $[-a,a]$. The density of eigenvalues can be derived from the resolvent by using \eqref{gdisc},
\be\label{rhosol} \rho(x) = \frac{1}{2\pi\la}\Bigl(x^{2}+1+\frac{a^{2}}{2}\Bigr)\sqrt{a^{2}-x^{2}}\quad\text{for}\ x\in [-a,a]\, .\ee
We then get straightforwardly exact formulas for any non-trivial planar correlator,
\be\label{corrsol} C_{2k} = \frac{(2k-1)!}{2^{k}3^{k+1}(k+2)!(k-1)!}\frac{1}{\la}\bigl(\sqrt{1+12\la} -1\bigr)^{k+1}\bigl(1+(1+k)\sqrt{1+12\la}\bigr)\, ,\ee
as well as for the free energy
\be\label{Fsol} F = -\frac{3}{8} + \frac{-1-36\la+(1+30\la)\sqrt{1+12\la}}{432\la^{2}} + \frac{1}{4}\ln\frac{1+6\la+\sqrt{1+12\la}}{2}\,\cdotp\ee

\subsubsection*{Perturbative and strong coupling expansions}

At small $\la$, \eqref{Fsol} yields the planar Feynman diagram expansion \eqref{Fpert} to any desired order, e.g.
\be\label{Fpertsol} F = \frac{\la}{2}-\frac{9\la^{2}}{8}+\frac{9\la^{3}}{2} - \frac{189\la^{4}}{8} + O(\la^{5})\, .\ee
It also yields a non-trivial strong coupling expansion of the form
\be\label{Fstrongsol} F = \frac{1}{2}\ln\sqrt{\la} + \frac{1}{8}\bigl(2\ln 3 - 3\bigr) + \frac{2}{3\sqrt{3}}\frac{1}{\sqrt{\la}}-\frac{1}{12\la} + O(\la^{-3/2})\, ,\ee
showing in particular that the strong coupling expansion parameter is $1/\sqrt{\la}$. The form of the perturbative expansion \eqref{Fpertsol} of course immediately follows from \eqref{ZNdef} and \eqref{SNdef} by rescaling $M\mapsto\sqrt{\la}M$ in the integral representation \eqref{ZNdef}. Interestingly, the form of the strong coupling expansion can also be understood directly from the integral representation, by rescaling $M\mapsto\la^{1/4}M$. The Gaussian integrals of the standard perturbation theory, with a weight $e^{-\frac{N}{2}\tr M^{2}}$, are then replaced by much more complicated-looking integrals with a weight $e^{-\frac{N}{4}\tr M^{4}}$ which are all automatically computed in the large $N$ limit by the solution \eqref{Fsol}.

\section{The probe brane action\label{s3Sec}} 
\subsection{Building the probe action\label{buildSec}}

To build the probe brane action, we carefully follow the steps explained in \cite{ferfund} (see in particular Section 6.3 in this reference) and use the results of \cite{ferequiv}. We focus on the case of a single probe brane.


\subsubsection*{Step one: equivariant gauge fixing}

The first step in the construction is to consider the model \eqref{SNdef}, but with $N+1$ colors instead of $N$, and to partially gauge-fix the gauge symmetry $\text{U}(N+1)$ down to $\uN\times\text{U}(1)$ \cite{ferequiv}. 

To do this, we introduce anticommuting ghosts $\eta_{a}$, $\bar\eta^{a}$ and anti-ghosts $\chi_{a}$, $\bar\chi^{a}$, together with commuting ``Lagrange multiplier'' variables $\La_{a}$ and $\bar\La^{a}$, all neutral under $\text{U}(1)$ but transforming in the fundamental or antifundamental representation of $\uN$ (note that these notations are in accord with the notations used in \cite{ferfund} but differ from the ones used in \cite{ferequiv}; for example, in Section 5 of \cite{ferequiv}, the ghosts and antighosts $\eta_{a},\bar\eta^{a},\chi_{a},\bar\chi^{a}$ are denoted as $\omega^{+},\omega^{-},\bar\omega^{+},\bar\omega^{-}$). The $(N+1)\times (N+1)$ matrix $M$ being decomposed as in \eqref{funddec}, we define an odd graded differential $\delta$ of ghost number one by
\begin{align}\label{deltadef0} \delta V^{a}_{\ b} & = i\bigl(\bar\eta^{a}w_{b}-\bar w^{a}\eta_{b}\bigr)\\\label{deltadef1}
\delta v &= i\bigl(\eta_{a}\bar w^{a}-w_{a}\bar\eta^{a}\bigr)\\\label{deltadef2}
\delta w_{a} &= i\bigl(\eta_{b}V^{b}_{\ a}-v\eta_{a}\bigr)\\\label{deltadef3}
\delta\bar w^{a} & = i\bigl(\bar\eta^{a}v - V^{a}_{\ b}\bar\eta^{b}\bigr)\\\label{deltadef4}
\delta\eta_{a} & = 0\, ,\
\delta\bar\eta^{a} =0\\\label{deltadef5}
\delta\chi_{a}&=-\La_{a}\, ,\
\delta\bar\chi^{a}=-\bar\La^{a}\\\label{deltadef6}
\delta\La_{a} & =
\chi_{b}\bar\eta^{b}\eta_{a}-\eta_{b}\bar\eta^{b}\chi_{a}\\\label{deltadef7}
\delta\bar\La^{a} & = \bar\chi^{a}\eta_{b}\bar\eta^{b}-\bar\eta^{a}\eta_{b}\bar\chi^{b}\, .
\end{align}
The square of the differential $\delta$ yields a $\uN\times\text{U}(1)$ gauge transformation and thus vanishes on $\uN\times\text{U}(1)$ invariants. For this reason, it is called an equivariant differential with respect to $\uN\times\text{U}(1)$. As explained in details in \cite{ferequiv}, $\delta$ is a version the usual Cartan differential used in the Cartan model of equivariant cohomology. It is thus directly related to the equivariant cohomology of $\text{U}(N+1)$ with respect to $\uN\times\text{U}(1)$, in the same way as the usual BRST differential used in the full gauge-fixing of a gauge group $G$ is related to the cohomology of $G$. 

We now consider the following equivariant gauge-fixing fermion,
\be\label{psidef} \psi = \bar\chi^{a}\Bigl(w_{a}-\frac{\xi}{2}\La_{a}\Bigr) + \chi_{a}\Bigl(\bar w^{a}-\frac{\xi}{2}\bar\La^{a}\Bigr)\, ,\ee
depending on an arbitrary gauge parameter $\xi\geq 0$. The associated gauge-fixing term that we must add to the action is 
\begin{multline}\label{deltapsi} -(N+1)\delta\psi = (N+1)\biggl[
-\xi\La_{a}\bar\La^{a}+\bar\La^{a}w_{a}+\La_{a}\bar w^{a}\\+ i\Bigl( \bigl(\bar\chi^{a}\eta_{b} + \bar\eta^{a}\chi_{b}\bigr)V^{b}_{\ a} + v\bigl(\eta_{a}\bar\chi^{a}+\chi_{a}\bar\eta^{a}\bigr)\Bigr)-\xi\Bigl( \bar\chi^{a}\chi_{b}\bar\eta^{b}\eta_{a} + \chi_{a}\bar\chi^{a}\eta_{b}\bar\eta^{b}\Bigr)\biggr]\, .
\end{multline}
The overall factor of $(N+1)$ has been inserted for later convenience.
The variables $\La_{a}$ and $\bar\La^{a}$ can be integrated out, which yields
\be\label{Lainout} \La_{a} = \frac{w_{a}}{\xi}\,\cvp\quad
\bar\La^{a} = \frac{\bar w^{a}}{\xi}\ee
and the final form of the equivariant gauge-fixing terms,
\begin{multline}\label{gfaction} \frac{1}{N+1}S_{\text{gauge-fixing}}[\xi] = 
\frac{1}{\xi}\bar w^{a}w_{a}+ i \eta_{a}\bigl(v\delta^{a}_{b}-V^{a}_{\ b} \bigr)\bar\chi^{b} + i\chi_{a}\bigl(v\delta^{a}_{b}-V^{a}_{\ b}\bigr)\bar\eta^{b}\\+ \xi\bigl(\eta_{a}\bar\eta^{a}\bar\chi^{b}\chi_{b} + \eta_{a}\bar\chi^{a}\chi_{b}\bar\eta^{b}\bigr)\, .
\end{multline}
The most remarkable property of this action is to include quartic ghost interactions. These interactions disappear in the limit $\xi\rightarrow 0$, in which case the term $\bar w^{a}w_{a}/\xi$ enforces the gauge conditions \eqref{Landau} strictly, but they are mandatory in general $\xi$-gauges. The crucial r\^ole played by the quartic ghost couplings to ensure the gauge invariance of the framework will be made very explicit later. 

\subsubsection*{Step two: the mixed action $S_{N,1}$}

By plugging the decomposition \eqref{funddec} into $S_{N+1}(M)$, we rewrite the total action as
\begin{multline}\label{stot0} S_{N+1}(M)+S_{\text{gauge-fixing}}[\xi](V,v,w,\bar w,\eta,\bar\eta,\chi,\bar\chi) = \\
S_{N}(V) + (N+1) S_{1}(v) + S_{N,1}[\xi](V,v,w,\bar w,\eta,\bar\eta,\chi,\bar\chi)\, .\end{multline}
The action $S_{N}(V)$ is the original matrix model action \eqref{SNdef} for the $N$ ``background'' branes, with matrix variable $V$. The action $S_{1}(v)$ is simply the classical potential for a single brane
\be\label{S1explit} S_{1}(v)=\frac{1}{\la}\Bigl(\frac{1}{2}v^{2}+\frac{1}{4}v^{4}\Bigr)\, .\ee
It comes with a factor $(N+1)$ in \eqref{stot0} because the action $S_{N+1}$ has the same factor in its definition. The action $S_{N,1}[\xi]$, which is gauge-dependent, describes the interactions between the ``background'' and the ``probe'' branes. It depends on ``off-diagonal'' variables, which include the ghosts and antighosts and which, in a string theory language, are associated with open strings stretched between the background and the probe branes. Explicitly, we find
\begin{multline}\label{SN1form} \frac{1}{N+1}S_{N,1}[\xi](V,v,w,\bar w,\eta,\bar\eta,\chi,\bar\chi) = \frac{1}{(N+1)\la}\tr_{N}\Bigl(\frac{1}{2}V^{2}+\frac{1}{4}V^{4}\Bigr)\\+\frac{1}{\la}
w_{a}\bigl( (1+v^{2})\delta^{a}_{b}+ v V^{a}_{\ b} + (V^{2})^{a}_{\ b}\bigr)\bar w^{b} + \frac{1}{2\la}\bigl(w_{a}\bar w^{a}\bigr)^{2} \\+
\frac{1}{\xi}w_{a}\bar w^{a}+ i \eta_{a}\bigl(v\delta^{a}_{b}-V^{a}_{\ b} \bigr)\bar\chi^{b} + i\chi_{a}\bigl(v\delta^{a}_{b}-V^{a}_{\ b}\bigr)\bar\eta^{b}\\+ \xi\bigl(\eta_{a}\bar\eta^{a}\bar\chi^{b}\chi_{b} + \eta_{a}\bar\chi^{a}\chi_{b}\bar\eta^{b}\bigr)\, .
\end{multline}
Note that the first term on the right-hand site of the above equation comes from the fact that the action $S_{N}$ is defined with a factor $N/\la$ whereas $S_{N+1}$ is defined with a factor $(N+1)/\la$.

Because the quartic ghost coupling is proportional to $\xi$, we see that the usual weak-coupling, perturbative regime (for a given $v$) corresponds in the above parameterization to having both $\la$ and $\xi$ small. This is not very convenient. It is much more natural to introduce a new gauge parameter $\zeta\geq 0$ defined by 
\be\label{zetadef} \xi = \zeta\la\, .\ee
The coupling $\la$ then recovers its usual loop counting parameter interpretation and the perturbative limit is $\la\rightarrow 0$ for any fixed $\zeta$. We shall always use the parameter $\zeta$ from now on.

\subsubsection*{Step three: the auxiliary fields}

We now rewrite the action $S_{N,1}[\zeta]$ in an equivalent form $\hat S_{N,1}[\zeta]$, such that the off-diagonal variables appear only quadratically. This is done by introducing five auxiliary fields $\mu,\gamma_{1},\gamma_{2},\gamma_{3},\gamma_{4}$ and choosing
\begin{multline}\label{hatSN1} \frac{1}{N+1}\hat S_{N,1}[\zeta](V,v,\mu,\gamma_{1},\gamma_{2},\gamma_{3},\gamma_{4},w,\bar w,\eta,\bar\eta,\chi,\bar\chi) =\frac{1}{(N+1)\la}\tr_{N}\Bigl(\frac{1}{2}V^{2}+\frac{1}{4}V^{4}\Bigr) \\+\frac{1}{\la}w_{a}\Bigl[ \bigl(1+\zeta^{-1}+v^{2}+\mu\bigr)\delta^{a}_{b}+ v V^{a}_{\ b} + (V^{2})^{a}_{\ b}\Bigr]\bar w^{b}-\frac{\mu^{2}}{2\la}\\ + i\eta_{a}\Bigl[\bigl(v+\gamma_{1}\bigr)\delta^{a}_{b}-V^{a}_{\ b}\Bigr]\bar\chi^{b} + i\chi_{a}\Bigl[\bigl(v+\gamma_{2}\bigr)\delta^{a}_{b}-V^{a}_{\ b}\Bigr]\bar\eta^{b}\\ + i \gamma_{3}\eta_{a}\bar\eta^{a} + i\gamma_{4}\chi_{a}\bar\chi^{a}+ \frac{\gamma_{1}\gamma_{2}-\gamma_{3}\gamma_{4}}{\zeta\la}\,\cdotp
\end{multline}
It is straightforward to check that, by integrating out $\mu,\gamma_{1},\gamma_{2},\gamma_{3},\gamma_{4}$, which yields
\begin{gather} \label{mudef} \mu = w_{a}\bar w^{a}\\
\label{gam12def}\gamma_{1}= -i\zeta\la\, \chi_{a}\bar\eta^{a}\, ,\
\gamma_{2}= -i\zeta\la\,\eta_{a}\bar\chi^{a}\\
\label{gam34def} \gamma_{3} = i\zeta\la\,\chi_{a}\bar\chi^{a}\, ,\
\gamma_{4}=i\zeta\la\,\eta_{a}\bar\eta^{a}\, ,
\end{gather}
the action \eqref{hatSN1} reduces to \eqref{SN1form}, as required.

\subsubsection*{Step four: the probe D-brane action}

The probe D-brane action $\mathscr A_{N,1}[\zeta]$ is defined by
\begin{multline}\label{AN1def} e^{-\mathscr A_{N,1}[\zeta](v,\mu,\gamma_{1},\gamma_{2},\gamma_{3},\gamma_{4})} = \frac{e^{-(N+1)S_{1}(v)}}{Z_{N}}
\la^{-2N-3}\zeta^{-N-2}
\int\!\d V\d w\d\bar w\d\eta\d\bar\eta\d\chi\d\bar\chi\\
e^{-S_{N}(V)-\hat S_{N,1}[\zeta](V,v,\mu,\gamma_{1},\gamma_{2},\gamma_{3},\gamma_{4},w,\bar w,\eta,\bar\eta,\chi,\bar\chi)}\, .
\end{multline}
The overall factors of $\la$ and $\zeta$ in the definition are inserted for convenience at this stage, see the next paragraph. In more general cases, a standard full gauge-fixing term for the $\uN$ group must also be inserted before doing the integral over the background brane fields $V$, but in the present zero-dimensional context this is not necessary because the volume of the gauge group is finite. The integrals over the off-diagonal fields $w,\bar w,\eta,\bar\eta,\chi,\bar\chi$ are Gaussian and can be performed exactly. We obtain in this way
\begin{multline}\label{AN1def2} e^{-\mathscr A_{N,1}[\zeta](v,\mu,\gamma_{1},\gamma_{2},\gamma_{3},\gamma_{4})} =\la^{-N-3}\zeta^{-N-2} e^{-\frac{N+1}{\la}(\frac{1}{2}v^{2}+\frac{1}{4}v^{4}-\frac{\mu^{2}}{2}+\frac{\gamma_{1}\gamma_{2}-\gamma_{3}\gamma_{4}}{\zeta})}\\
\Bigl\langle e^{-\frac{1}{\la}\tr_{N}(\frac{1}{2}V^{2}+\frac{1}{4}V^{4})}\frac{\Delta_{\text{gh}}}{\Delta_{w,\bar w}}\Bigr\rangle\, ,
\end{multline}
where the expectation value is taken in the original $\uN$ matrix model with action $S_{N}(V)$ and $\Delta_{\text{gh}}$ and $\Delta_{w,\bar w}$ are $\uN$ invariant Pfaffian and determinant operators given by
\begin{align}\label{Deltagh} \Delta_{\text{gh}} &= \text{Pf}\!
\begin{pmatrix} 0 & \gamma_{3} \mathbb I_{N}& 0 & (v+\gamma_{1})\mathbb I_{N}-V\\
-\gamma_{3}\mathbb I_{N} & 0 & -(v+\gamma_{2})\mathbb I_{N}+V & 0\\
0 & (v+\gamma_{2})\mathbb I_{N} - V & 0 & \gamma_{4}\mathbb I_{N}\\
-(v+\gamma_{1})\mathbb I_{N}+V & 0 & -\gamma_{4}\mathbb I_{N} & 0
\end{pmatrix}
\\\label{Deltaww} \Delta_{w,\bar w} &= \det\Bigl[ \bigl( 1+\zeta^{-1}+v^{2}+\mu\bigr)\mathbb I_{N}+v V + V^{2}\Bigr]\, .
\end{align}

In the leading large $N$ limit we're interested in, the above formulas simplify. First, the probe brane action $\mathscr A_{N,1}[\zeta]\simeq N A[\zeta]$ can be treated classically in this limit. This means that various variables can be integrated out by simply solving their classical equations of motion. Because of ghost number conservation and the fact that a bosonic global symmetry cannot be spontaneously broken in zero dimension, we know that solving
\be\label{Agam34eq} \frac{\partial A[\zeta]}{\partial\gamma_{3}}= \frac{\partial A[\zeta]}{\partial\gamma_{4}}=0\ee
must yield the on-shell values
\be\label{gamma34star} \gamma_{3}^{*}=\gamma_{4}^{*} = 0\, .\ee
Similarly, since $A[\zeta]$ depends on $\gamma_{1}$ and $\gamma_{2}$ symmetrically, we can integrate out one of these variables and set
\be\label{gamm120}\gamma_{1}=\gamma_{2}=\gamma\, .\ee
The Pfaffian in \eqref{Deltagh} then reduces to the square of a determinant,
\be\label{Deltagh2} \Delta_{\text{gh}} = \Bigl(\det\bigl( (v+\gamma)\mathbb I_{N} - V\bigr)\Bigr)^{2}\, .\ee
Moreover, because of the large $N$ factorization of gauge-invariant correlation functions, the expectation value in \eqref{AN1def2} can be much simplified and evaluated by using the master field \eqref{rhosol}. Overall, we obtain 
\begin{multline}\label{Aform} A[\zeta](v,\mu,\gamma) = \frac{1}{\la}\Bigl(\frac{1}{2}v^{2}+\frac{1}{4}v^{4}-\frac{\mu^{2}}{2}+\frac{\gamma^{2}}{\zeta}\Bigr) +\ln\bigl(\zeta\la\bigr)-\frac{3}{2}
+\frac{1}{\la}\Bigl(\frac{1}{2}C_{2}(\la)+\frac{1}{4}C_{4}(\la)\Bigr)\\
+\int_{-a}^{a}\rho(x)\Bigl[\ln\bigl(1+\zeta^{-1}+\mu+v^{2}+v x+x^{2}\bigr)-\ln \bigl(v+\gamma-x\bigr)^{2}\Bigr]\d x \, .
\end{multline}
In the ``Landau'' gauge $\zeta=0$, this formula simplifies considerably. The variables $\mu$ and $\gamma$ can be integrated out by setting $\mu=\gamma=0$ and we get
\be\label{Azetazeroform} A[0](v) = \frac{1}{\la}\Bigl(\frac{1}{2}v^{2}+\frac{1}{4}v^{4}\Bigr) 
+\frac{1}{\la}\Bigl(\frac{1}{2}C_{2}(\la)+\frac{1}{4}C_{4}(\la)\Bigr)-\frac{3}{2}
-\int_{-a}^{a}\!\d x\,\rho(x)\ln \frac{(v-x)^{2}}{\la} \, \cdotp
\ee
The formula also simplifies in the opposite $\zeta=\infty$ limit. In this limit, the ghost condensate $\gamma$ scales as $\sqrt{\la\zeta}$ and the $\gamma$-dependence in \eqref{Aform} is simply given by $\gamma^{2}/(\la\zeta)-\ln\gamma^{2}$. The ghost condensate can thus be integrated out and we get
\begin{multline}\label{Azetainfiniteform} A[\infty](v,\mu) = \frac{1}{\la}\Bigl(\frac{1}{2}v^{2}+\frac{1}{4}v^{4}-\frac{\mu^{2}}{2}\Bigr)-\frac{1}{2}
+\frac{1}{\la}\Bigl(\frac{1}{2}C_{2}(\la)+\frac{1}{4}C_{4}(\la)\Bigr)\\
+\int_{-a}^{a}\rho(x)\ln\bigl(1+\mu+v^{2}+v x+x^{2}\bigr)\,\d x \, .
\end{multline}

Note that the constants appearing in \eqref{Aform} are required for the fundamental identity \eqref{ZNZN1A} discussed in the next subsection to be valid. As explained after Eq.\ \eqref{SN1form}, the terms involving $C_{2}$ and $C_{4}$ simply come from the definition of the 't~Hooft coupling in the theories with $N$ and $N+1$ colors. The term $\ln(\zeta\la)$ comes from the overall factors in the path integral \eqref{AN1def}. These are akin to the overall $\la^{-N^{2}/2}$ factor in the definition \eqref{ZNdef} and are easy to derive from first principles, see the Appendix. The additional constant $-3/2$ can be obtained by looking at the $\la\rightarrow 0$ limit and using the normalization $Z_{N}(\lambda=0)=1$, or equivalently $F(\lambda=0)=0$, consistently with the definition \eqref{ZNdef}. This will be checked explicitly in the next section.

\subsection{Basic properties of the probe action\label{baspropSec}}

The probe brane action $\mathscr A_{N,1}[\zeta]$ defined by \eqref{AN1def2}, including the overall factors of $\la$ and $\zeta$, automatically satisfies \cite{ferfund,ferequiv}
\be\label{ZNZN1A} \frac{Z_{N+1}}{Z_{N}} = \int\!\d v\d\mu\d\gamma_{1}\d\gamma_{2}\d\gamma_{3}\d\gamma_{4}\, e^{-\mathscr A_{N,1}[\zeta](v,\mu,\gamma_{1},\gamma_{2},\gamma_{3},\gamma_{4})}\, .\ee
In order the make the present paper as self-contained as possible, we derive this basic identity from first principles in the Appendix, using the path integral approach to the equivariant gauge-fixing procedure explained in Section 4.3 of Ref.\ \cite{ferequiv}. As discussed in \cite{ferfund}, in the large $N$ limit,
\be\label{largeNZZFF} \ln Z_{N}= -N^{2}F + O(N^{0})\, ,\quad \mathscr A_{N,1}[\zeta]= NA[\zeta] + O(N^{0})\, .\ee
Eq.\ \eqref{ZNZN1A} is thus equivalent to the fundamental relation \eqref{AFrel} between the on-shell D-brane action and the free energy, here in the case $K=1$ on which we have been focusing,
\be\label{onshellA} F = \frac{1}{2}A[\zeta]^{*}\, .\ee
In particular, since the free energy depends only on the coupling $\la$ but of course not on the gauge-fixing parameter $\zeta$, this identity implies that the minimal value of $A[\zeta]$ must also be independent of $\zeta$. This property is not manifest on the explicit formula \eqref{Aform}, but we shall be able to check it explicitly in the next Section.

\section{The exact solution\label{s4Sec}}
\subsection[]{The solution in the gauge $\zeta=0$\label{Seczetazero}}
\begin{figure}
\centerline{\includegraphics[width=6in]{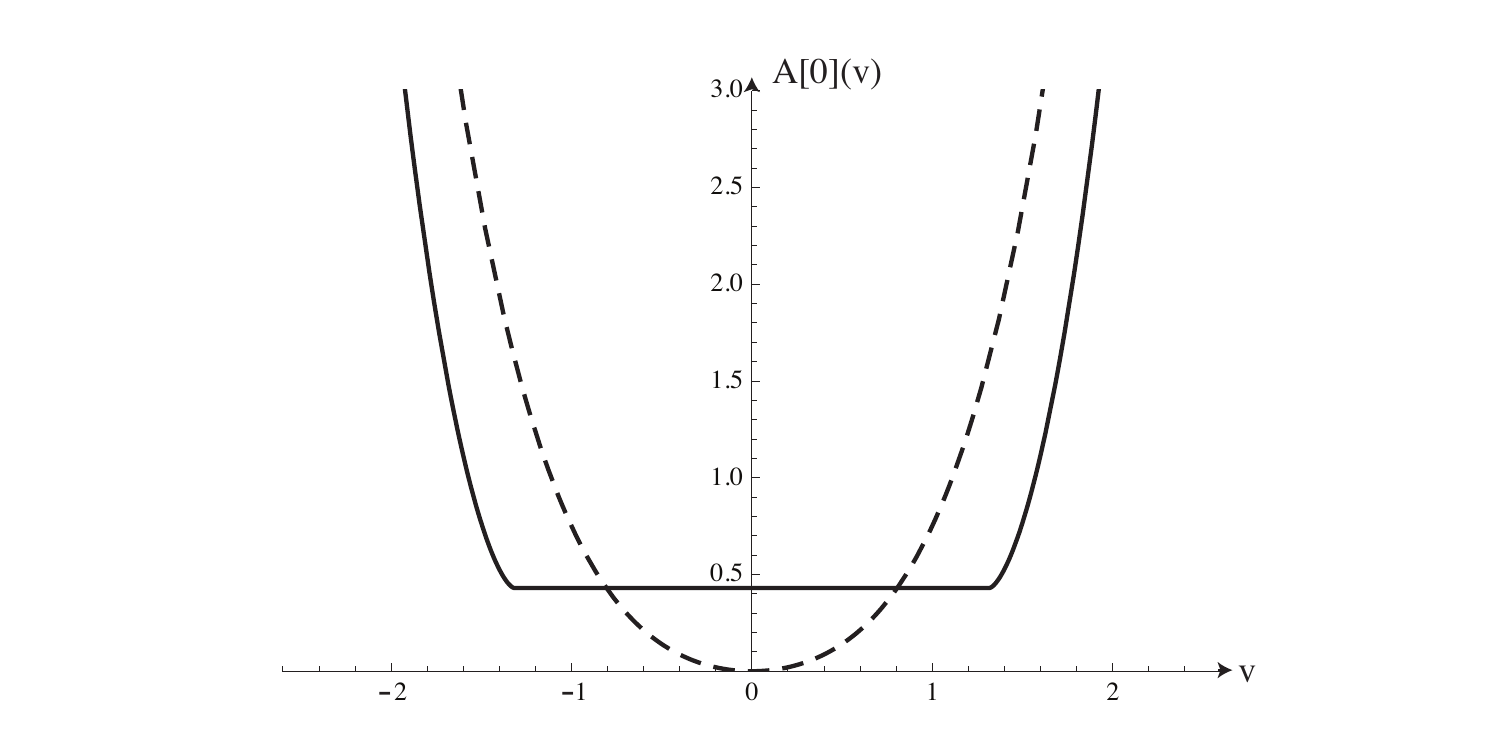}}
\caption{The classical (dashed line) and quantum (plain line) probe action as a function of $v$, in the gauge $\zeta=0$, for a typical strong 't~Hooft's coupling $\la=1$. The value of the quantum action at its minimum is twice the free energy of the matrix model, as dictated by \eqref{onshellA}. In this gauge, the $v$-dependence of the quantum action coincides with the well-known quantum potential for an eigenvalue, which is the sum of the classical potential with a repulsive, Coulomb-like potential due to the background eigenvalues.
\label{fig1}}
\end{figure}

Let us start by discussing the D-brane probe action \eqref{Azetazeroform} in the Laudau gauge $\zeta=0$. Since in this case the off-diagonal components $w$ and $\bar w$ of the matrix $M$ in the decomposition \eqref{funddec} strictly vanish, the variable $v$ is an eigenvalue. The action $A[0](v)$ is then simply the traditional effective quantum potential for an eigenvalue in the presence of all the others: the first term in \eqref{Azetazeroform} is the classical potential that the eigenvalue feels, and the last term is the traditional repulsive Coulomb force produced by all the other eigenvalues.

\subsubsection*{The exact off-shell D-brane action}

Evaluating this effective quantum potential is a standard exercise. If we take the derivative of \eqref{Azetazeroform} with respect to $v$, we find
\be\label{derA0v} \frac{\d A[0](v)}{\d v} = \frac{v+v^{3}}{\la} - 2\text{P}\!\!\int_{-a}^{a}\frac{\rho(x)}{v-x}\,\cdotp\ee
The saddle point equation \eqref{saddlerho} implies that
\be\label{derA0v2} \frac{\d A[0](v)}{\d v} = 0 \quad \text{if}\ v\in [-a,a]\, .\ee
If $|v|>a$, the integral \eqref{derA0v} can be easily done and yields
\be\label{derA0v3} \frac{\d A[0](v)}{\d v} = \frac{1}{\la}\biggl[v^{3}+\Bigl(1+\frac{a^{2}}{2}\Bigr)v\biggr]\sqrt{1-\frac{a^{2}}{v^{2}}} \quad \text{if}\ |v|>a\, .\ee
Integrating \eqref{derA0v2} and \eqref{derA0v3} and using the continuity of the action at $v=\pm a$ (which follows from the integral representation \eqref{Azetazeroform}), we get
\be\label{A0exact1} A[0](v) = 
\frac{1}{\la}\biggl[\frac{v^{4}}{4} + \Bigl(1+\frac{a^{2}}{4}\Bigr)\frac{v^{2}}{2}\biggr]\sqrt{1-\frac{a^{2}}{v^{2}}} - 2\ln\frac{|v|+\sqrt{v^{2}-a^{2}}}{a} + 2f(\la) \quad \text{if $|v|>a$}\ee
and
\be\label{A0exact2} A[0](v) = 2f(\la)\quad \text{if $v\in [-a,a]$,}\ee
for some $v$-independent constant $f(\la)$. This constant can be fixed by looking at the large $v$ asymptotics of $A[0](v)$. From \eqref{Azetazeroform}, we find
\be\label{A0asymp} A[0](v) = \frac{1}{\la}\Bigl(\frac{1}{2}v^{2}+\frac{1}{4}v^{4}\Bigr) 
+\frac{1}{\la}\Bigl(\frac{1}{2}C_{2}(\la)+\frac{1}{4}C_{4}(\la)\Bigr)-\frac{3}{2}-\ln\frac{v^{2}}{\la}\ee
and, comparing with the expansion derived from \eqref{A0exact1} by using \eqref{asol}, we get
\be\label{fFrel} f(\la) = F(\la)\, ,\ee
where $F$ is the free energy \eqref{Fsol}. A graph of the D-brane action $A[0](v)$, for the typical strong coupling $\la=1$, is plotted in Fig.\ \ref{fig1}.

\subsubsection*{The exact on-shell D-brane action}

To go on-shell, we have to find the absolute minimum of $A[0]$. From \eqref{derA0v2} and \eqref{derA0v3}, we derive that this minimum is obtained for any value $v^{*}\in [-a,a]$. Using \eqref{A0exact2} and \eqref{fFrel}, the corresponding critical value is found to coincide with twice the free energy, consistently with \eqref{onshellA}. 

\subsubsection*{Bulk locality and the Gribov problem}

The quantum potential felt by the probe eigenvalue has the well-known but remarkable property of being exactly flat for $v\in [-a,a]$. In the D-brane/holographic interpretation, this means that the probe D-brane feels no force in the ``deep IR'' region of the bulk geometry. This may seem very surprising. No-force conditions are usually associated with very special cancellations coming from supersymmetry, but obviously this cannot be the explanation here.

In the present case, the flatness of the action should be rather interpreted as coming from an extreme manifestation of the non-locality of the deep bulk geometry. In particular, it must be a strongly gauge-dependent phenomenon \cite{ferfund}. More precisely, we are going to explain that it is directly related to the Gribov problem in the gauge $\zeta=0$.

In this gauge, the off-diagonal components $w$ and $\bar w$ of the matrix $M$ in \eqref{funddec} are set to zero, which identifies $v$ as being one eigenvalue of the matrix $M$. Since $M$ has $N+1$ eigenvalues, which are permuted by an $\text{S}_{N+1}$ subgroup of $\text{U}(N+1)$, this produces $N+1$ Gribov copies, according to which precise eigenvalue of $M$ the variable $v$ should be identified to. This Gribov ambiguity is harmless in the path integral treatment since it simply produces an overall factor of $N+1$. However, it has deep consequences on the bulk spacetime. The interval $[-a,a]$ represents the support of the density of eigenvalues of the matrix $V$ in \eqref{funddec} and is thus filled with the $N$ eigenvalues of the matrix $M$ which are different from $v$ (the ``background branes''). Since all the $N+1$ eigenvalues are indistinguishable due to the $\text{S}_{N+1}$ gauge symmetry, \emph{all the spacetime points in the interval $[-a,a]$ should be identified} from the point of view of the probe. 

In other words, in the bulk, the gauge group acts as the group $\text{S}_{\infty}$ of all bijections, smooth or not, from the interval $[-a,a]$ onto itself. This group can be viewed as the large $N$ limit of $\text{S}_{N+1}$ and is more general than an ordinary group of smooth diffeomorphisms. In particular, moving the probe brane anywhere in the interval $[-a,a]$ is a gauge transformation! The flatness of the action $A[0]$ in thus a consequence of gauge invariance. One may say that the space interval $[-a,a]$ has really collapsed to a single point.

The above discussion, in our very simple model, suggests more generally that the Gribov problem could have rather deep consequences from the point of view of the holographic spacetime and that the problem of locality in the bulk is even more stringent than what is suggested by the gauging of ordinary smooth diffeomorphisms.

\subsection[]{The solution in a general $\zeta$-gauge}

In the general $\zeta$-gauge, the D-brane probe action $A[\zeta]$ given in \eqref{Aform} is a natural function of three variables $v$, $\mu$ and $\gamma$. In this sense, the bulk holographic geometry is naturally three-dimensional. As explained in \cite{ferfund}, one could, even more generally, integrate in an arbitrary number of variables of the form
\be\label{integratein} w_{a}(V^{p})^{a}_{\ b}\bar w^{b}\, ,\quad
\eta_{a}(V^{p})^{a}_{\ b}\bar \chi^{b}\, \quad \text{etc.,}\ee
for any positive integer $p$ (this could of course be done in the gauge $\zeta=0$ as well). These variables correspond to the excited string modes living on the D-brane probe. All the actions obtained in this way are strictly equivalent. Physically, only their on-shell values is relevant and these all match. Of course, in higher-dimensional models, when there is a separation of mass scales between the modes, it is natural to work with only the lowest mass modes and approximate the resulting action in a derivative expansion.

In the present zero-dimensional example, we mainly focus on the natural formulation with the three variables $v$, $\mu$ and $\gamma$. We also consider the possibility of integrating out two variables out of the three, to get an action interpolating smoothly between the general $\zeta$-gauges and the solution in the gauge $\zeta=0$ described in the preceding subsection. It turns out that the off-shell description in terms of $v$ only becomes singular if $\zeta$ is greater than a critical value. One may then use the combination
\be\label{vpdef} v' = v+\gamma\ee
of $v$ and the ghost condensate $\gamma$ to get a well-defined off-shell description for all values of $\zeta$.

\subsubsection*{The exact off-shell D-brane actions}

Let us introduce 
\be\label{rdef} r = r_{1} + i r_{2} = -\frac{v}{2} + \frac{i}{2}\sqrt{3 v^{2} + 4\bigl(1+\zeta^{-1}+\mu\bigr)}\, ,\ee
such that
\be\label{idwithr} |x-r|^{2}=(x-r)(x-\bar r) = x^{2} + v x + v^{2} + \mu +\zeta^{-1} + 1\, .\ee
One can then evaluate the integrals
\begin{align}\label{int1} \int_{-a}^{a}\frac{\rho(x)}{(x-r)(x-\bar r)}\,\d x &= -\frac{\im g(r)}{\la r_{2}}\\
\label{int2} \int_{-a}^{a}\frac{x\rho(x)}{(x-r)(x-\bar r)}\,\d x &= -\frac{\im (rg(r))}{\la r_{2}}\,\cvp
\end{align}
by using \eqref{gdisc} and the residue theorem. This yields explicit formulas for the partial derivatives of the probe brane action \eqref{Aform},
\begin{align}\label{der1} \frac{\partial A[\zeta]}{\partial\gamma} & =
\begin{cases}\displaystyle\vphantom{\Bigg|}\frac{2\gamma}{\zeta\la}-\frac{(v+\gamma)^{3}+v+\gamma}{\la} & \text{if $v+\gamma\in [\-a,a]$}\\\displaystyle
\vphantom{\Bigg|}\frac{2\gamma}{\zeta\la} - \frac{2g(v+\gamma)}{\la} & \text{if $|v+\gamma|>a$}
\end{cases}
\\\label{der2}
\frac{\partial A[\zeta]}{\partial\mu} & =-\frac{\mu}{\la} -\frac{\im g(r)}{\la r_{2}}\\
\label{der3}\frac{\partial A[\zeta]}{\partial v} & =
\begin{cases}\displaystyle\vphantom{\Bigg|}\frac{v^{3}-(v+\gamma)^{3}-\gamma}{\la}-\frac{2 v\im g(r)+\im (rg(r))}{\la r_{2}}& \text{if $v+\gamma\in [\-a,a]$}\\
\displaystyle\vphantom{\Bigg|} \frac{v+v^{3}-2g(v+\gamma)}{\la}-\frac{2 v\im g(r)+\im (rg(r))}{\la r_{2}}& \text{if $|v+\gamma|>a$.}
\end{cases}
\end{align}
These derivatives can be straightforwardly integrated to obtain an explicit formula for $A[\zeta]$, but this is not particularly illuminating. One can also easily eliminate $\gamma$ and $\mu$ by setting to zero \eqref{der1} and \eqref{der2} and work with an action $A[\zeta](v)$ or $A[\zeta](v')$, where $v'$ is defined in \eqref{vpdef}.

If we insist on using $v$ for all values of $\zeta$, we actually run into a difficulty. The problem is that the equation
\be\label{der10}\frac{\partial A[\zeta]}{\partial\gamma} = 0\ee
cannot be used to find $\gamma$ as a well-defined function of $v$ for all values of $\zeta$. Indeed, in terms of $v'$, \eqref{der10} is easily solved,
\be\label{gammavsvp} \gamma = 
\begin{cases} \frac{1}{2}\zeta\bigl( v'^{3}+v'\bigr) & \text{if $v'\in [-a,a]$}\\
\zeta g(v') & \text{if $|v'|>a$.}
\end{cases}\ee
However, the relation between $v'$ and $v$ takes the form
\be\label{vpvsv} v= f(v')\, ,\ee
where the function $f$ is defined by
\be\label{deff} f(v') = 
\begin{cases} 
v'-\frac{\zeta}{2}\bigl(v'+v'^{3}\bigr) & \text{if $v'\in [-a,a]$}\\
v'-\zeta g(v') & \text{if $|v'|>a$.}
\end{cases}\ee
A completely elementary analysis, using the explicit formula \eqref{gsol} for $g$, then shows that $f$ is a strictly monotonic function, and is thus invertible, as long as
\be\label{zetac} \zeta \leq \zeta_{\text{c}} = \frac{1}{\sqrt{1+12\la} - 1/2}\,\cdotp\ee
If $\zeta>\zeta_{\text c}$, $f$ is no longer invertible. The inverse $f^{-1}$ and the action $A[\zeta](v)$ then become ill-defined, multivalued functions, in some range of the variable $v$. Note that this singular behavior only occurs if one insists in using the variable $v$ off-shell. Indeed, the on-shell value $v^{*}$ of $v$ always lies in the domain where $A[\zeta](v)$ is well-defined. The problem actually disappears altogether when $\zeta\rightarrow\infty$. In this limit, $v^{*}\rightarrow 0$ and the range of values of $v$ for which $A[\zeta](v)$ is ill-defined vanishes.

\begin{figure}
\centerline{\includegraphics[width=6in]{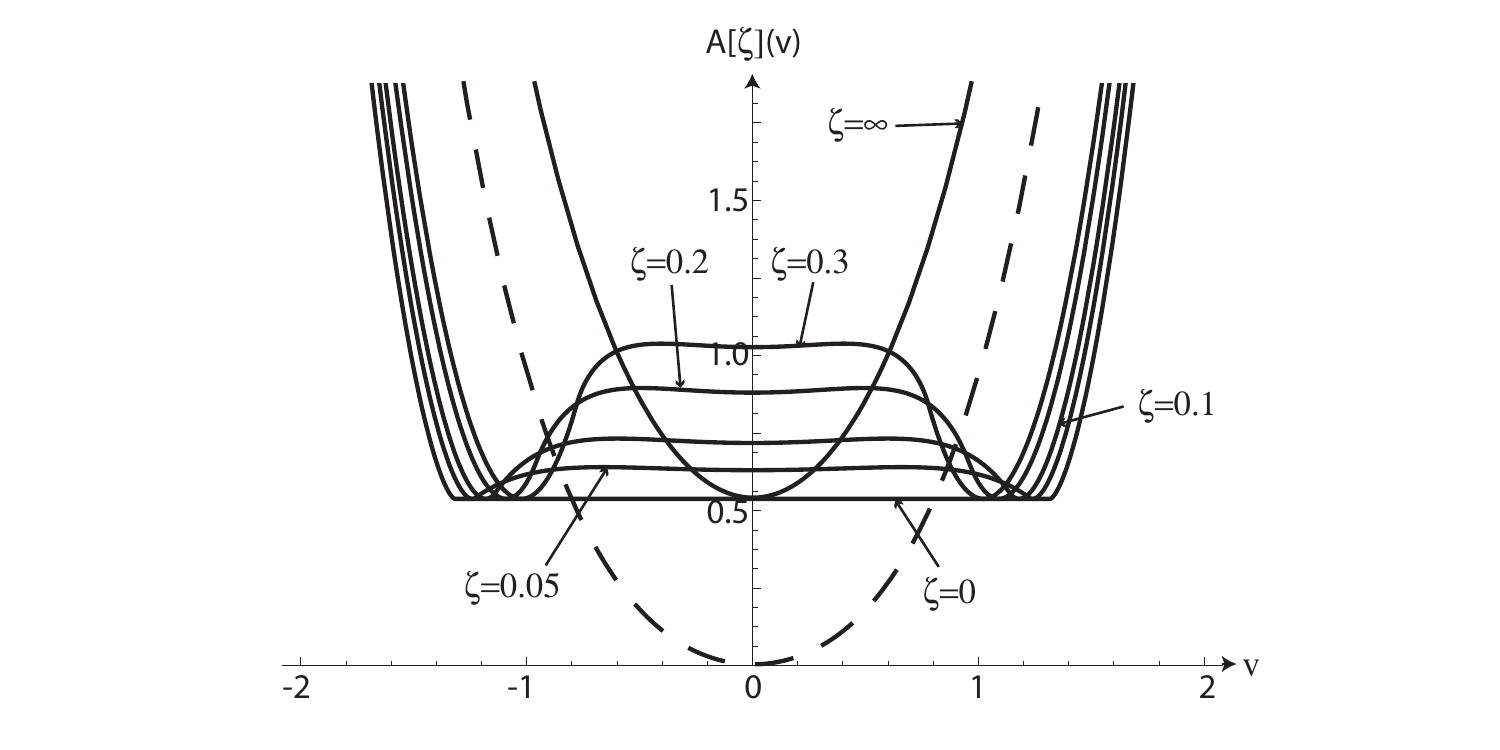}}
\caption{Exact probe actions $A[\zeta]$ as a function of $v$, for $\la=1$ and various values of the gauge parameter $0\leq\zeta\leq\zeta_{\text c}$ and $\zeta=\infty$. The dashed line represents the classical action.\label{fig2}}
\end{figure}
\begin{figure}
\centerline{\includegraphics[width=6in]{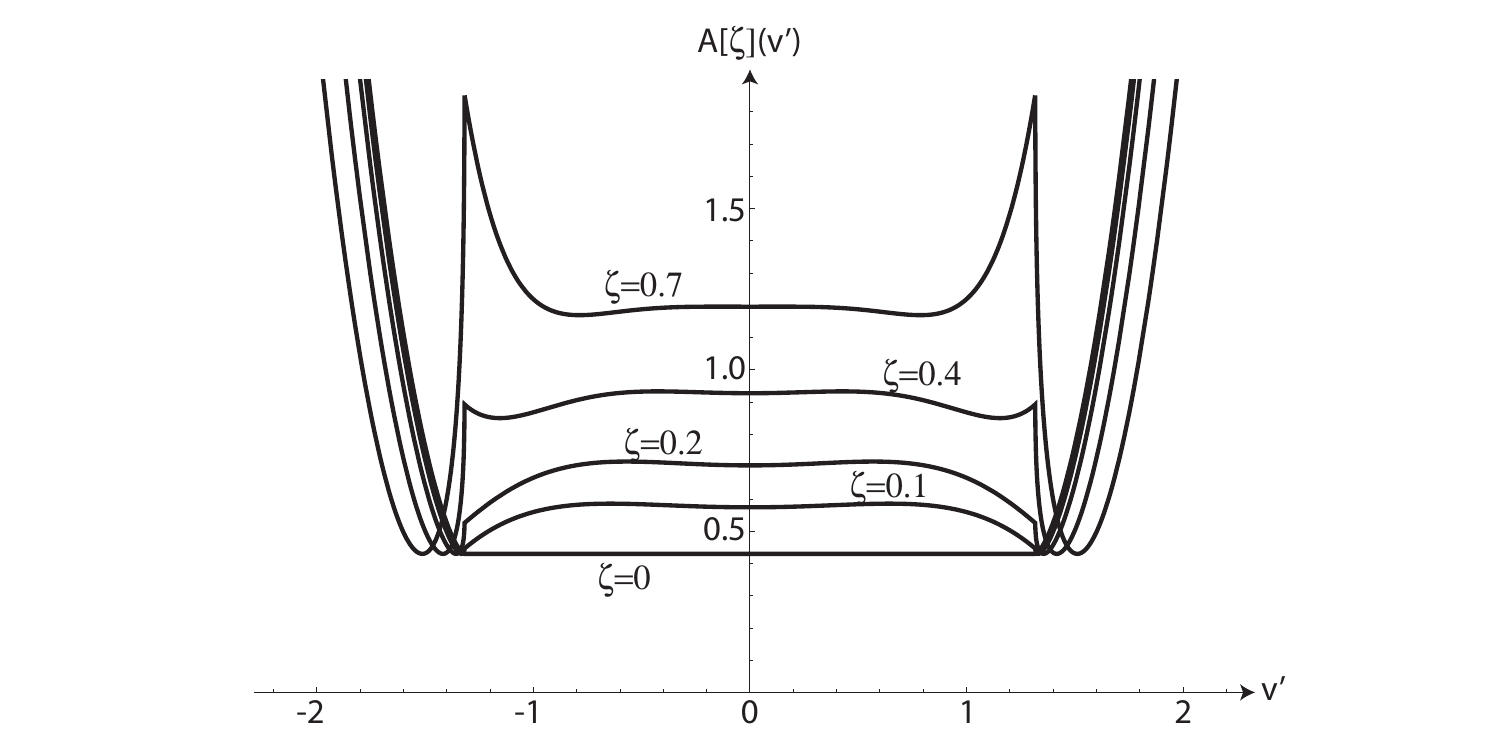}}
\caption{Exact probe actions $A[\zeta]$ as a function of $v'$ defined in \eqref{vpdef}, for $\la=1$ and various values of the gauge parameter $\zeta$.\label{fig3}}
\end{figure}

The resulting actions $A[\zeta]$ are depicted in Fig.\ \ref{fig2} and in Fig.\ \ref{fig3} for various values of $\zeta$ and for a typical strong  't~Hooft's coupling $\la=1$. 

\subsubsection*{Gauge dependence, bulk locality and diffeomorphisms}

All these actions look pretty much the same in the large $v$ or $v'$ ``UV'' region, but they differ considerably in the deep ``IR'' region, for $v,v'\sim a$. This is consistent with the general discussion in \cite{ferfund}: the gauge-dependence is negligible in the UV but becomes very strong deep in the bulk. This strong gauge-dependence of the holographic description in the deep bulk is directly related to the fact that local bulk physics is not observable \cite{ferfund}.

A natural guess is that the gauge ambiguity in the probe action should be related to the ambiguities in the bulk description coming from the closed string gauge symmetries, like diffeomorphism invariance \cite{insightbrane,ferfund}. In our present very simple set-up, we only have the reparameterization ambiguity. We are thus led to ask whether we can always find, for any given choices of gauge parameters $\zeta$ and $\zeta'$, a diffeomorphism $v\mapsto \varphi(v)$ such that
\be\label{diffeoonA} A[\zeta']\bigl(\varphi(v)\bigr)\overset{?}{=}A[\zeta](v)\, .\ee
Note that, if true, this would automatically ensure the gauge-independence of the on-shell physics, since the minimal value of a function cannot depend on the way you parameterize its argument. 

\begin{figure}
\centerline{\includegraphics[width=6in]{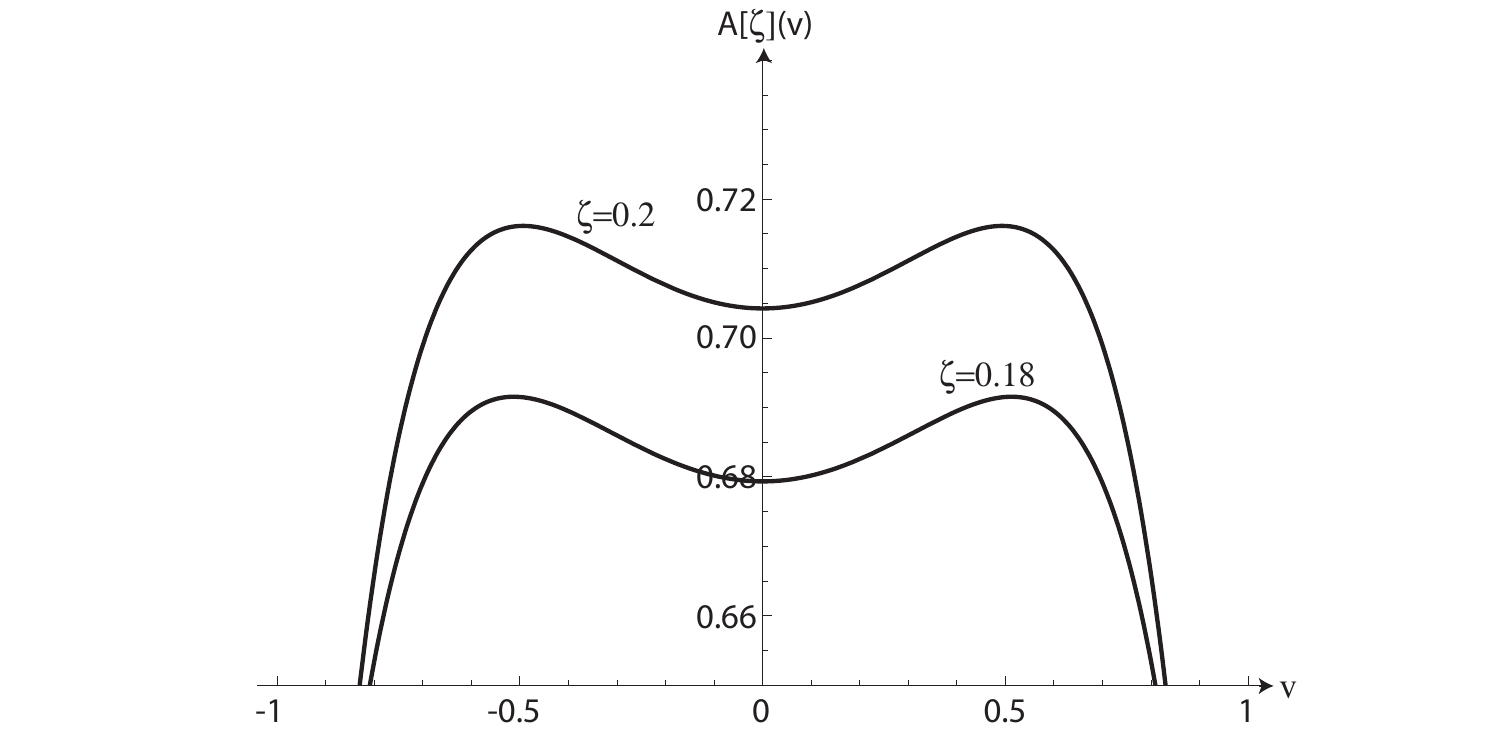}}
\caption{Probe actions for two different values of the gauge parameter, zoomed in the deep bulk region. The actions have three critical points in this region, with $\zeta$-dependent critical values. This shows that the actions cannot be related by a smooth change of parameterization.\label{fig4}}
\end{figure}

However, and this may come as a surprise, the relation \eqref{diffeoonA} cannot be valid deep in the bulk geometry. A very simple way to understand this point is to note that \eqref{diffeoonA} would imply that the set of \emph{all} the critical values of $A[\zeta]$ and $A[\zeta']$, not only the minimum, should match. However, this is not true. In the deep bulk region, these actions can have three critical points, with $\zeta$-dependent critical values, as the graphs depicted in Fig.\ \ref{fig4} clearly show. It is actually straightforward to check analytically that the action \eqref{Aform} always have a critical point at $v=\gamma=0$ and some fixed non-zero value of $\mu$, with a $\zeta$-dependent critical value. Other ways to see that the relation between the various actions cannot be simply a smooth reparameterization is to note that the variable $v$ becomes singular in the deep bulk when $\zeta$ is greater than the critical value \eqref{zetac}; or to note that the action at $\zeta=0$ is flat in the deep bulk, whereas the actions at $\zeta>0$ are not.

\begin{figure}
\centerline{\includegraphics[width=6in]{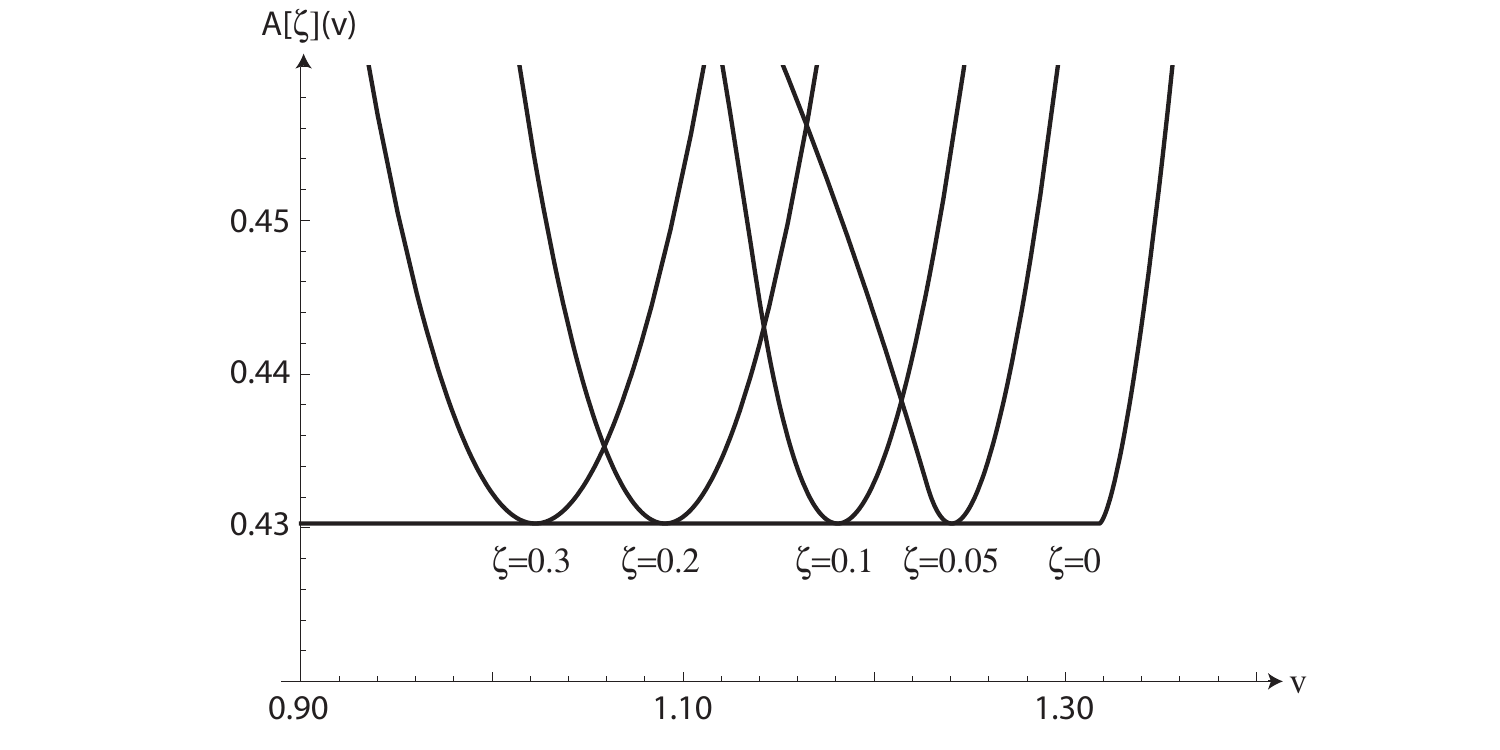}}
\caption{The same probe actions as in Fig.\ \ref{fig2} zoomed in the vicinity of their absolute minima. All these minima coincide with twice the value of the free energy $2F(1)\simeq 0.4303$.\label{fig5}}
\end{figure}

\begin{figure}
\centerline{\includegraphics[width=6in]{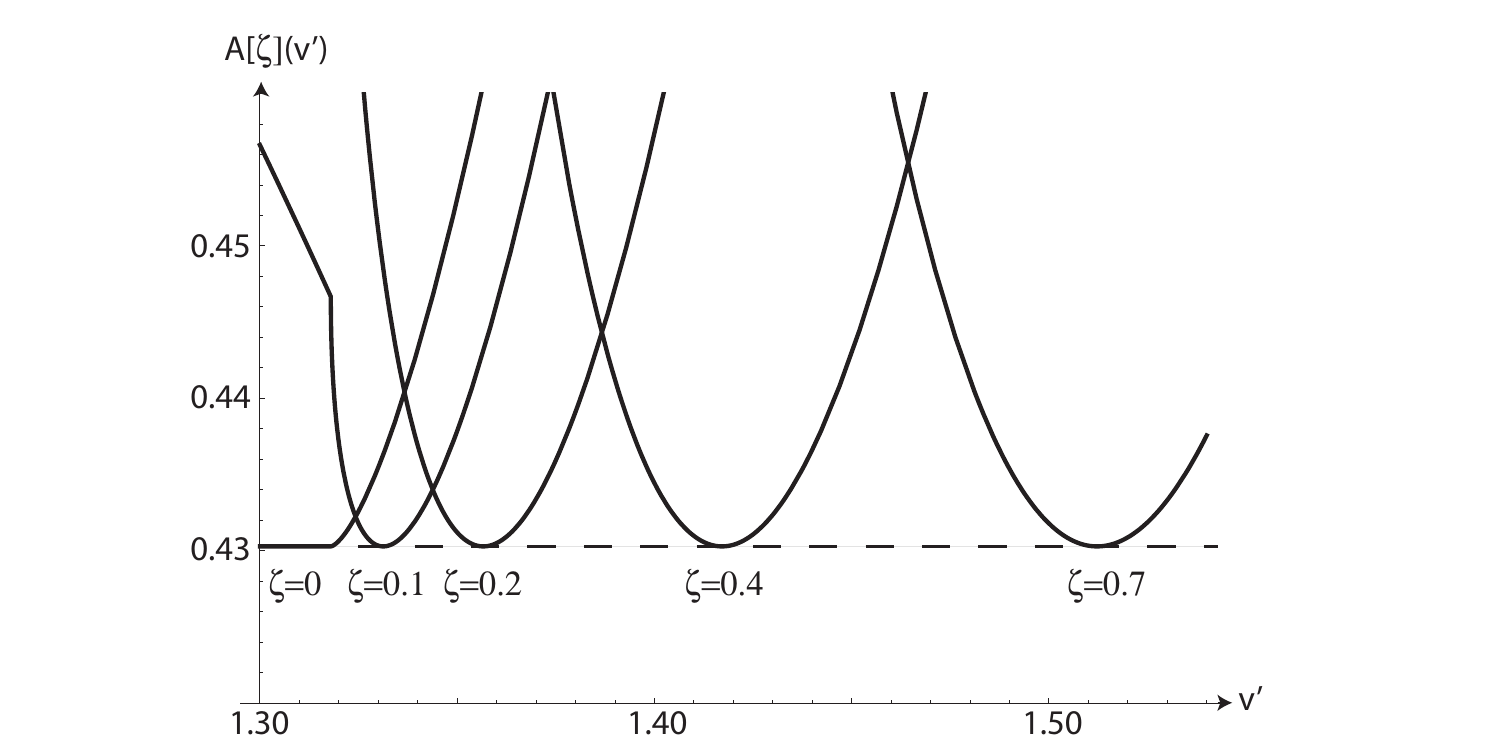}}
\caption{The same probe actions as in Fig.\ \ref{fig3} zoomed in the vicinity of their absolute minima. All these minima coincide with twice the value of the free energy $2F(1)\simeq 0.4303$ (dashed line).\label{fig6}}
\end{figure}

Still, the physical critical value of the actions must be $\zeta$-independent and coincide with twice the free energy. Without \eqref{diffeoonA}, this property might seem quite miraculous, but it is ensured by the overall consistency of the formalism. It is also manifest on the numerical graphs, in particular Figs.\ \ref{fig5} and \ref{fig6}. We shall provide a completely explicit proof below. 

One consequence of the gauge-independence of the physical critical value is that, if we limit ourselves to a region excluding the deep bulk interval between the points $-v^{*}[\zeta]$ and $v^{*}[\zeta]$ where the minimum of the action is reached, then there do exist diffeomorphisms relating all the different actions. It's in the deep bulk region $-v^{*}[\zeta]<v<v^{*}[\zeta]$ that the gauge-dependence is more general than simply diffeomorphisms.

Of course, the present model is a very simplified example of holography and one must be careful when drawing conclusions or even intuitions for more interesting cases in higher dimensions. However, we do believe that the effects of the gauge-dependence on the deep bulk description of probes will be dramatic in all cases. In particular, it might shed a very interesting new light on spacetimes containing black holes and on the physics of a probe ``crossing the horizon.''

\subsubsection*{On-shell gauge-independence, equivariant BRST spontaneous symmetry breaking and ghost condensation}


The critical points ($v^{*},\mu^{*},\gamma^{*})$ of the action $A[\zeta]$ are found by setting to zero the partial derivatives \eqref{der1}--\eqref{der3}. From \eqref{Aform}, we can then compute the derivative of the critical values $A[\zeta]^{*}$ with respect to $\zeta$,
\be\label{derAzeta}\frac{\partial A^{*}[\zeta]}{\partial\zeta} = 
-\frac{\gamma^{2}}{\la\zeta^{2}} + \frac{1}{\zeta} - \frac{1}{\zeta^{2}}\int_{-a}^{a}\frac{\rho(x)}{1+\zeta^{-1}+\mu^{*}+v^{*2}+v^{*} x + x^{2}}\,\d x\, .\ee
Using the equation of motion $\partial A[\zeta]/\partial\mu=0$ for $\mu$, this expression can be simplified to
\be\label{derAzeta2}\frac{\partial A^{*}[\zeta]}{\partial\zeta}  = \frac{\la\zeta-\mu^{*}-\gamma^{*2}}{\la\zeta^{2}}\,\cdotp\ee
We thus see that the on-shell gauge-independence is equivalent to the existence of a critical point for which the relation
\be\label{BRSTsym} \mu^{*}+\gamma^{*2}=\la\zeta\ee
holds.

\begin{figure}
\centerline{\includegraphics[width=6in]{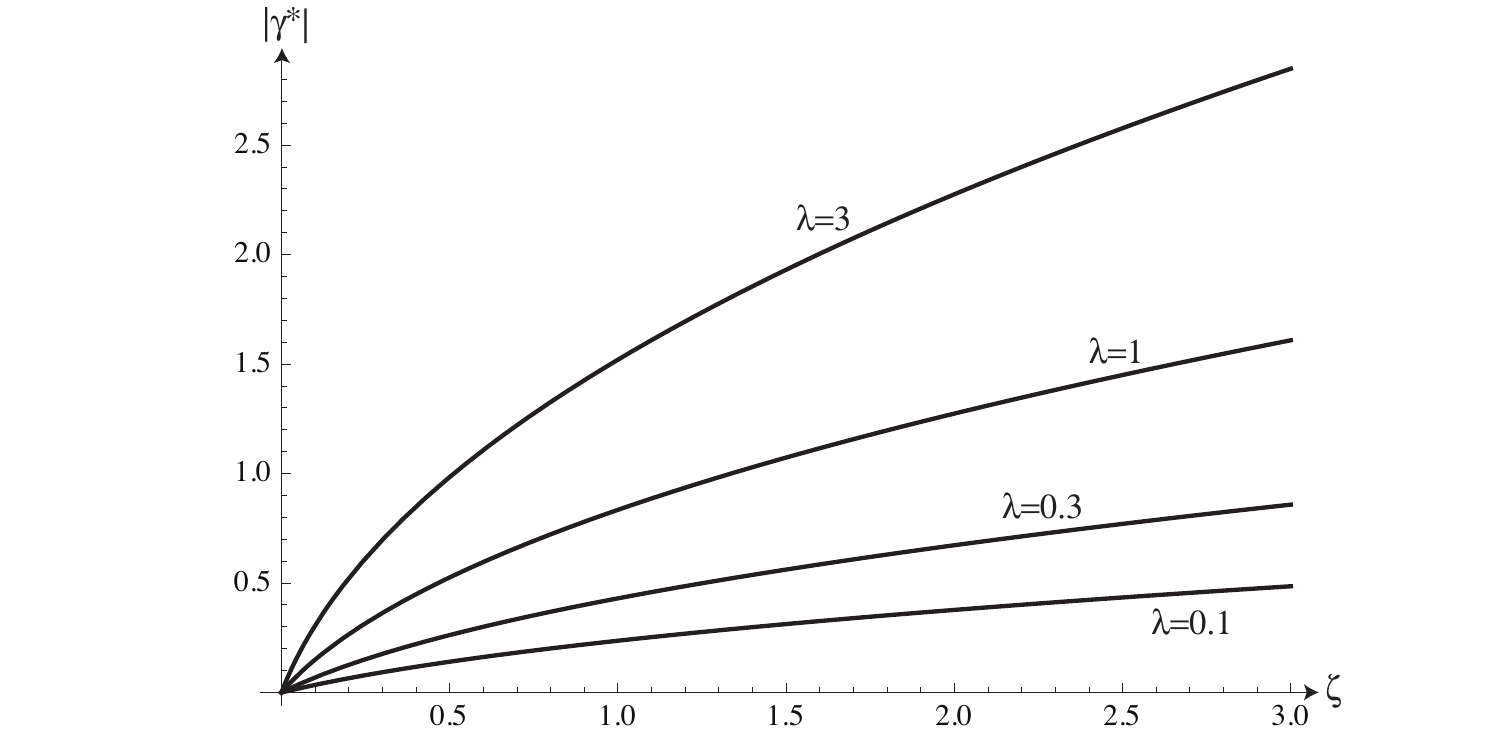}}
\caption{\label{fig7}The ghost condensate $|\gamma^{*}|$ as a function of the gauge fixing parameter $\zeta$, for various values of the coupling $\la$. Ghost condensation is a characteristic feature of the physical critical point of the probe actions.}
\end{figure}

The condition \eqref{BRSTsym} has a very interesting consequence. Let us compute the expectation value of the equivariant BRST variation of $\chi_{a}\bar\La^{a}$ and $\La_{a}\bar\chi^{a}$. Using \eqref{Lainout}, \eqref{mudef}--\eqref{gam34def}, \eqref{gamma34star} and \eqref{gamm120}, we get from \eqref{deltadef5}--\eqref{deltadef7} that
\be\label{vevBRS} \bigl\langle\delta (\chi_{a}\bar\La^{a})\bigr\rangle = 
\bigl\langle\delta (\La_{a}\bar\chi^{a})\bigr\rangle = -\frac{\mu^{*}+ \gamma^{*2}}{(\la\zeta)^{2}}\, \cdotp\ee
This result is quite remarkable. It shows that the on-shell gauge-independence, which is equivalent to \eqref{BRSTsym}, implies the spontaneous breakdown of the equivariant BRST symmetry,
\be\label{BRSTbroken} \bigl\langle\delta (\chi_{a}\bar\La^{a})\bigr\rangle = 
\bigl\langle\delta (\La_{a}\bar\chi^{a})\bigr\rangle = -\frac{1}{\la\zeta}\not = 0\, .\ee
We believe that this feature is generic and will occur in much more complicated and realistic models. It is clearly related to the existence of the quartic ghost terms in the equivariant ghost Lagrangian. A closely related phenomenon, also made possible by the presence of the quartic ghost terms, is ghost condensation. Using the system of Eqs.\ \eqref{S1eq1}, \eqref{S2eq2} that we derive in the next paragraph, one can easily check that
\be\label{ghostcond} \gamma^{*} = -i\zeta\la\bigl\langle \chi_{a}\bar\eta^{a}\bigr\rangle = -i\zeta\la\bigl\langle \eta_{a}\bar\chi^{a}\bigr\rangle\not = 0\ee
at the physical critical point where \eqref{BRSTsym} is valid. The ghost condensate is depicted in Fig.\ \ref{fig7}.

\subsubsection*{Explicit proof of the on-shell gauge-independence}

As we have repeatedly emphasized, we know a priori that the relation \eqref{BRSTsym} must hold, since gauge invariance is automatically built in the formalism and follows from \eqref{onshellA}. However, it is rather non-trivial from the point of view of the equations of motion. We wish now to provide a direct derivation, which we believe is quite instructive and illustrates well a fundamental and non-trivial property of the equivariant gauge-fixing procedure.\footnote{Note that, in practice, taking for granted gauge invariance, we could use \eqref{BRSTsym} to greatly simplify the analysis of the equations of motion.} 

In principle, we have two things to prove: first, the on-shell gauge-independence, which is equivalent to \eqref{BRSTsym}; second, that the corresponding $\zeta$-independent critical value coincides with twice the free energy. This second point actually follows immediately from the first and the result of Section \ref{Seczetazero} at $\zeta=0$. We thus focus on \eqref{BRSTsym}.  

We start from the equations of motion
\be\label{eom2} \frac{\partial A[\zeta]}{\partial\gamma} = \frac{\partial A[\zeta]}{\partial\mu} =\frac{\partial A[\zeta]}{\partial v} =0\, ,\ee
where the partial derivatives are given explicitly in \eqref{der1}--\eqref{der3}. There are two cases to consider, depending on whether $v'=v+\gamma\in[-a,a]$ or not. These two cases can be studied along the same lines. The case $v'\in [-a,a]$ yields critical points with $\zeta$-dependent critical values. This was checked  numerically in Fig.\ \ref{fig4} and can be trivially checked analytically as well. We thus focus on the other case, $|v'|>a$. From \eqref{der1}--\eqref{der3}, we then find that \eqref{eom2} implies
\begin{align}\label{eq1} g(v+\gamma) & = \frac{\gamma}{\zeta}\\\label{eq2}
\frac{\im g(r)}{r_{2}} & = -\mu\\\label{eq3} \frac{\im(r g(r))}{r_{2}} & =
v+v^{3}-\frac{2\gamma}{\zeta} + 2\mu v\, ,
\end{align}
where $r$, $r_{1}=\re r$ and $r_{2}=\im r$ are defined by \eqref{rdef}. The conditions \eqref{eq2} and \eqref{eq3} allow to find both $\re g(r)$ and $\im g(r)$, which yields
\be\label{gr} 
g(r) = v+v^{3}+\frac{3}{2}\mu v-\frac{2\gamma}{\zeta} - i r_{2}\mu\, .\ee
Since the resolvent \eqref{gsol} satisfies the degree two equation
\be\label{deg2res} \Bigl(g(z)-\frac{z^{3}+z}{2}\Bigr)^{2} = \frac{1}{4}\Bigl(z^{2}+1+\frac{a^{2}}{2}\Bigr)^{2}\bigl(z^{2}-a^{2}\bigr)\, ,\ee
Eq.\ \eqref{gr} implies one complex algebraic equation involving $r_{1}$, $r_{2}$ and $\gamma$. The real part of this equation yields
\begin{multline}\label{real1} 4\gamma^{2} - 2r_{1}\bigl(6 + (1+r_{1}^{2}-3r_{2}^{2})\zeta\bigr)\gamma +3\zeta r_{1}^{4}+\zeta r_{2}^{4}+
\frac{\zeta^{2}}{16} a^{2}\bigl(2+a^{2}\bigr)^{2}\\-\bigl(
1+\zeta+\la\zeta^{2})r_{2}^{2} + \bigl(9+(3-12 r_{2}^{2}+\la\zeta)\zeta\bigr)r_{1}^{2} = 0
\end{multline}
and the imaginary part yields
\be\label{im1}
\bigl(3r_{1}^{2}-r_{2}^{2}+1+2\zeta^{-1}\bigr)\gamma = r_{1}\bigl(
3\zeta^{-1}+2+5r_{1}^{2}-3r_{2}^{2}+\la\zeta\bigr)\, .\ee
Similarly, by using \eqref{deg2res}, we find that \eqref{eq1} implies 
\begin{multline}\label{neweq} \gamma^{4} - 6 r_{1}\gamma^{3} +\bigl(1+12 r_{1}^{2}-\zeta^{-1}-\la\zeta\bigr)\gamma^{2} - 2\bigl(1+4r_{1}^{2}-2\la\zeta\bigr) r_{1}\gamma\\-\frac{\zeta}{16}a^{2}\bigl(2+a^{2}\bigr)^{2}-4\la\zeta r_{1}^{2} = 0\, .
\end{multline}
Eqs.\ \eqref{real1}, \eqref{im1} and \eqref{neweq} provide three algebraic equations for the three variables $\gamma$, $r_{1}$ and $r_{2}$ (or equivalently $\gamma$, $\mu$ and $v$). These algebraic equations are equivalent to the equations of motion in the case $|v'|>a$.

Let us introduce the new variable
\be\label{defZ} Z = \mu+\gamma^{2}=r_{2}^{2}-3r_{1}^{2}-1-\zeta^{-1}\ee
and express $r_{2}^{2}$ in terms of $Z$, $r_{1}$ and $\gamma$ in \eqref{real1}--\eqref{neweq}. We get
\begin{align}\label{E1eq} 
\begin{split}&
\gamma^{4}-6 r_{1}\gamma^{3}+\bigl(6r_{1}^{2}+3\zeta^{-1}-1+\la\zeta-2Z\bigr)\gamma^{2} + 2\bigl(2+8r_{1}^{2}+3Z-3\zeta^{-1}\bigr)r_{1}\gamma
\\ &\hskip 2cm  +Z^{2} +\bigl(\zeta^{-1}+1-6r_{1}^{2}-\la\zeta\bigr)Z 
+\frac{\zeta}{16}a^{2}\bigl(2+a^{2}\bigr)^{2}\\
&\hskip 6cm - 24 r_{1}^{4}-(1+\zeta)\la -2(3+\zeta\la)r_{1}^{2} = 0
\end{split}\\\label{E2eq} &
\gamma^{3} - 3r_{1}\gamma^{2}+\bigl(\zeta^{-1}-Z\bigr)\gamma
+\bigl(1+4r_{1}^{2}+3Z-\zeta\la\bigr)r_{1} = 0\\\label{E3eq}
\begin{split}&
\gamma^{4}-6 r_{1}\gamma^{3}+\bigl(1+12 r_{1}^{2}-\zeta^{-1}-\zeta\la\bigr)\gamma^{2}-2\bigl(1+4r_{1}^{2}-2\zeta\la\bigr)r_{1}\gamma\\
&\hskip 6 cm - \frac{\zeta}{16}a^{2}\bigl(2+a^{2}\bigr)^{2} - 4\zeta\la r_{1}^{2} = 0\, .
\end{split}
\end{align}
We now massage this system of equation a little bit. First, subtracting \eqref{E3eq} from \eqref{E1eq}, we get a degree two equation in $\gamma$,
\begin{multline}\label{deg2gamma} 
2\bigl(1+3r_{1}^{2}+Z-2\zeta^{-1}-\la\zeta\bigr)\gamma^{2} +
2\bigl(-3-12r_{1}^{2}-3Z+3\zeta^{-1}+2\la\zeta\bigr)r_{1}\gamma-Z^{2}\\
+\bigl(6r_{1}^{2}-1-\zeta^{-1}+\la\zeta\bigr)Z-\frac{\zeta}{8}a^{2}\bigl(2+a^{2}\bigr)^{2}+24 r_{1}^{4}+(1+\zeta)\la-2(\la\zeta-3)r_{1}^{2}=0\, .
\end{multline}
Second, by using \eqref{E2eq}, we find a degree three equation from \eqref{E1eq},
\begin{multline}\label{deg3gamma} 3r_{1}\gamma^{3}-\bigl(1+12r_{1}^{2}
+Z-2\zeta^{-1}-\zeta\la\bigr)\gamma^{2} - \bigl(5\zeta\la - 3 - 12r_{1}^{2}-3Z\bigr)r_{1}\gamma\\
+\frac{\zeta}{16}a^{2}\bigl(2+a^{2}\bigr)^{2}+ 4\zeta\la r_{1}^{2} = 0\, .
\end{multline}
Third, by subtracting $3r_{1}$ times Eqs.\ \eqref{E2eq} to \eqref{deg3gamma}, we get a new degree two equation in $\gamma$,
\begin{multline}\label{def2gamma} \bigl(1+3 r_{1}^{2}+Z-2\zeta^{-1}-\zeta\la\bigr)\gamma^{2}+\bigl(-3-12r_{1}^{2}-6Z+3\zeta^{-1}+5\zeta\la\bigr)r_{1}\gamma\\ + \bigl(3 + 12r_{1}^{2}+9Z-7\zeta\la\bigr)r_{1}^{2}
-\frac{\zeta}{16}a^{2}\bigl(2+a^{2}\bigr)^{2} = 0\, .
\end{multline}
Finally, subtracting twice this equation to \eqref{deg2gamma}, we get the very simple result
\be\label{lasteq} \bigl(Z-\zeta\la\bigr)\bigl(6r_{1}\gamma-\zeta^{-1}-1-12r_{1}^{2}-Z\bigr) = 0\, .\ee
It is straightforward to check that the branch of the solution with $Z\not = \zeta\la$ is imaginary and must be discarded. We have thus derived
\be\label{Zcond} Z^{*} = \mu^{*}+\gamma^{*2} = \la\zeta\, ,\ee
which is the relation \eqref{BRSTsym} we wished to prove. The other two independent variables $r_{1}=-v/2$ and $\gamma$ are then solution of the equations
\begin{align}\label{S1eq1} &\gamma^{3} -3r_{1}\gamma^{2}+\bigl(\zeta^{-1}-\zeta\la\bigr)\gamma +\bigl(1+4r_{1}^{2}+2\zeta\la\bigr)r_{1} = 0\\
\label{S2eq2}
\begin{split} & \bigl(1+3r_{1}^{2}-2\zeta^{-1}\bigr)\gamma^{2}+
\bigl(3\zeta^{-1}-\zeta\la-3-12r_{1}^{2}\bigr)r_{1}\gamma
\\ & \hskip 4cm
+ \bigl(3+12r_{1}^{2}+2\zeta\la\bigr)r_{1}^{2} -\frac{\zeta}{16}a^{2}\bigl(2+a^{2}\bigr)^{2} = 0\, ,
\end{split}
\end{align}
which can be solved straightforwardly and always have two real solutions $(\gamma^{*},v^{*})$ and $(-\gamma^{*},-v^{*})$.

\section{The bare bubble approximation\label{s5Sec}}

The bare bubble approximation is a new non-perturbative approximation to matrix gauge theories which is naturally suggested by the probe D-brane formalism \cite{ferfund}. It is in principle tractable in any model, including the pure Yang-Mills theory in four dimensions. Its main interest is that it does not depend on the existence of a small parameter and is thus fundamentally non-perturbative. Our aim in this Section is to illustrate the main properties of the approximation and to assess its reliability by comparing the ``bare bubble'' free energy in the matrix model with the exact result.    

\subsection{Brief description of the approximation}
\begin{figure}
\centerline{\includegraphics[width=6in]{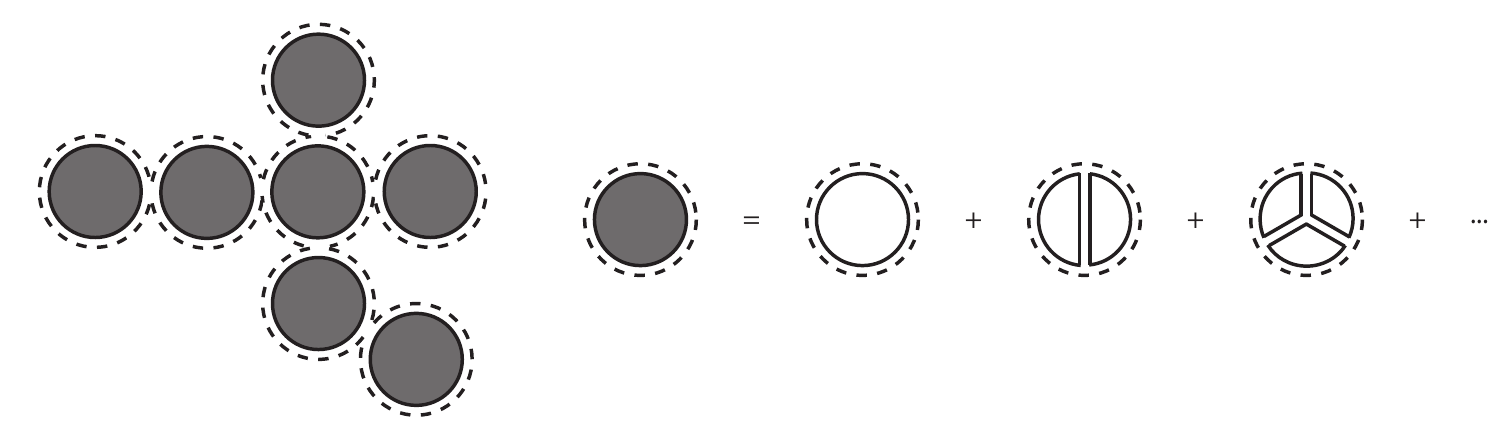}}
\caption{\label{fig8}A typical planar diagram contributing to the large $N$ probe D-brane action (left inset). The dashed lines are associated with probe brane indices (which, for a single probe, can take only one value) and the plain lines to background brane indices (which take $N$ values). The diagrams have a tree-like structure made of dressed bubbles, which themselves pick contributions from an infinite set of planar diagrams (right inset).}
\end{figure}
\begin{figure}
\centerline{\includegraphics[width=6in]{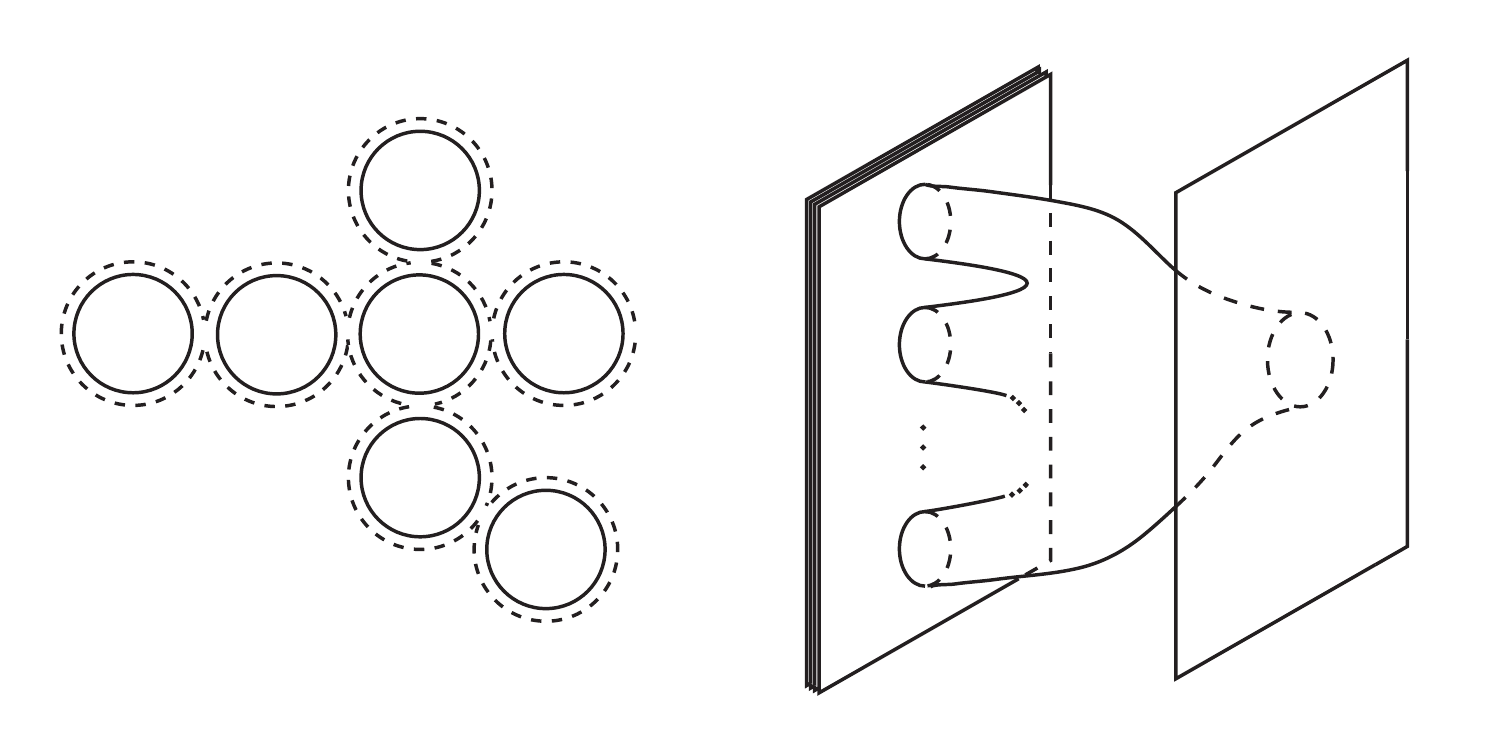}}
\caption{\label{fig9}A typical planar diagram contributing in the bare bubble approximation (left inset) and its worldsheet representation (right inset). For a bare bubble diagram, the number of loops, or equivalently the number of worldsheet boundaries on the background branes, coincide with the number of bubbles in the diagram.}
\end{figure}

As explained in \cite{ferfund}, the leading large $N$ probe D-brane action can be computed by summing an infinite series of planar diagrams having a mixed vector/matrix model structure, as depicted in Fig.\ \ref{fig8}. Each diagram can be represented in the form of a tree. The vertices in the tree take the form of bubbles which can be dressed by the quantum interactions on the background branes. The bare bubble approximation amounts to keeping only the simplest possible ``bare'' bubbles, as in Fig.\ \ref{fig9}.

At first sight, this may seem to be a rather gross approximation to make. Keeping only the bare bubbles means that we are treating the fields living on the background branes classically. In particular, all the probe brane degrees of freedom of the form \eqref{integratein}, for $p\geq 1$, are frozen and thus the open string theory on the probe branes is truncated. Yet, the interactions between the probe and the background branes coming from the couplings between the probe brane fields and the open strings stretched between the probe and the background are treated exactly. This produces an infinite series of planar diagrams, with an arbitrary number of loops. The approximation is thus potentially able to capture some features of the strongly coupled physics.

The bare bubble approximation is not a systematic expansion in a small parameter and this is its main interest. It can be used a priori for any matrix gauge theory. It is very different from the usual weak coupling expansion, which is also an expansion in the topology of the worldsheet diagrams. Actually, all the worldsheet topologies contributing to the large $N$ probe brane action are taken into account in the bare bubble approximation, see Fig.\ \ref{fig9}. Instead of a truncation in topologies, the approximation is rather a truncation of the worldsheet moduli space for each given topology, keeping only the Feynman diagrams for which the number of loops and the number of bubbles coincide.

There is a price to pay for not having a small parameter. If $\tilde A[\zeta]$ denotes the D-brane action computed in the bare bubble approximation, then there is no reason for the on-shell approximate action $\tilde A[\zeta]^{*}$ to be $\zeta$-independent. This means that strict gauge invariance must be lost. However, \emph{this is not an inconsistency of the approximation}. This is actually a required feature of \emph{any} approximation scheme in gauge theory that does not rely on the existence of a small expansion parameter.\footnote{For approximations that correspond to an expansion in a small parameter, the gauge invariance of the exact result, for all values of the expansion parameter, automatically implies the gauge invariance of the expansion to any given order.}  What we really have is an infinite family of approximations, one for each equivariant gauge choice. The gauge choice may then be adjusted to improve the approximation.

Of course, if the approximation is to be useful, all the gauge choices within a reasonable range should yield similar results and capture correctly the physics. In our case, this means that the bare bubble free energy
\be\label{Fbarebubble} \tilde F(\la;\zeta) = \frac{1}{2}\tilde A[\zeta]^{*}\ee
should not depend too violently on $\zeta$ and should match with a reasonable accuracy the exact free energy $F(\la)$ for a wide range of values of $\zeta$. In particular, this must be true for values of $\zeta$ that are singled out by some ``good'' properties one may wish to impose on the solution. For example, we may fix $\zeta$ by imposing that the approximate free energy is two-loop exact (by construction, it is always one-loop exact). As we shall see below, this yields an excellent approximation to the exact result all the way from the weakly to the very strongly coupled regime!

\subsection{The bare bubble free energy}

From what we have explained in the previous subsection, the probe brane action in the bare bubble approximation is obtained from the exact formula \eqref{Aform} by neglecting the quantum fluctuations of the background brane variables. This amounts to setting the background brane density of eigenvalue to its classical value $\rho(x)=\delta(x)$,
\be\label{Aformbb} \tilde A[\zeta](v,\mu,\gamma) = \frac{1}{\la}\Bigl(\frac{1}{2}v^{2}+\frac{1}{4}v^{4}-\frac{\mu^{2}}{2}+\frac{\gamma^{2}}{\zeta}\Bigr) +\ln\la 
+\ln\bigl(1+\zeta^{-1}+\mu+v^{2}\bigr)-\ln (v+\gamma)^{2}+c(\zeta) \, .
\ee
We are allowing for a coupling- and field-independent additive constant $c(\zeta)$, which will be adjusted to ensure the proper normalization of the bare bubble partition function, 
\be\label{normaliza} \lim_{\la\rightarrow 0}\tilde F(\la;\zeta) = \frac{1}{2}\lim_{\la\rightarrow 0}\tilde A[\zeta]^{*} = 0\, .\ee
The equations of motion derived from \eqref{Aformbb} are equivalent to the conditions
\begin{align}
\label{eombb1} &\gamma = \frac{\zeta}{2}v\bigl(1+2\mu+v^{2}\bigr)\\
\label{eombb2} & \gamma(\gamma+v) = \zeta\la\\
\label{eombb3} & \mu\bigl(1+\zeta^{-1}+\mu + v^{2}\bigr) = \la\, .
\end{align}
When $\la\rightarrow 0$, this yields $\mu\simeq \zeta\la/(1+\zeta)$, $\gamma^{2}\simeq\zeta^{2}\la/(2+\zeta)$ and $v^{2}\simeq 4\la/(2+\zeta)$. Plugging these asymptotics into \eqref{Aformbb} and using \eqref{normaliza}, we get
\be\label{czeta} c(\zeta) = \ln\frac{\zeta(\zeta+2)}{\zeta+1} - 1\, .\ee

Eqs.\ \eqref{eombb1}--\eqref{eombb3} are straightforward to solve and explicit formulas for the on-shell values of $v$, $\mu$ and $\gamma$ and for the bare bubble free energy $\tilde F(\la;\zeta)$ can be found. For example, at $\zeta=0$, we have
\be\label{tildeFzero} \tilde F(\la;0) = -\frac{1}{4}+\frac{\sqrt{1+8\la}-1}{16\la}-\frac{1}{2}\ln\frac{\sqrt{1+8\la}-1}{4\la}\,\cvp\ee
with weak and strong coupling expansions of the form
\begin{align}\label{weakbb0} \tilde F(\la;0) & = \frac{\la}{2}-\la^{2} + O\bigl(\la^{3}\bigr)\\\label{strongbb0}
\tilde F(\la;0) & = \frac{1}{2}\ln\sqrt{\la} + \frac{1}{4}\bigl(\ln 2 -1\bigr)+\frac{1}{2\sqrt{2}}\frac{1}{\sqrt{\la}} + O\bigl(1/\la\bigr)\, .
\end{align}
In the opposite $\zeta\rightarrow\infty$ limit, we find the amusing relation
\be\label{tildeFinf} \tilde F(\la;\infty) = \tilde F(\la/2;0)\, .\ee
For $\zeta>0$, the explicit formulas are very complicated and not particularly illuminating, so we shall refrain to copy them here. We simply indicate the weak and strong coupling expansions,
\begin{align}\label{smalllabb} \tilde F(\la;\zeta) &= \frac{8+\zeta(\zeta+2)^{2}(\zeta+8)}{4(\zeta+1)^{2}(\zeta+2)^{2}}\,\la + O\bigl(\la^{2}\bigr)\\\label{stronglabb} \tilde F(\la;\zeta) &=\frac{1}{2}\ln\sqrt{\la} -\frac{1}{4} + \frac{1}{2}\ln\frac{2+\zeta}{1+\zeta} + \frac{1}{2\sqrt{\la}}+O\bigl(1/\la\bigr)\, .
\end{align}
Let us note that the $\la\rightarrow\infty$ and $\zeta\rightarrow 0$ limits do not commute: the subleading terms in \eqref{strongbb0} are not obtained by taking the $\zeta\rightarrow 0$ limit of the subleading terms in \eqref{stronglabb}. This comes from the fact that, when $\la\rightarrow\infty$, $v\sim 1/\sqrt{\zeta}$ when $\zeta>0$ whereas $v\sim (2\la)^{1/4}$ when $\zeta = 0$.

The main properties of the bare bubble free energy $\tilde F(\la;\zeta)$ can be summarized as follows.

\noindent (i) The qualitative features of both the weak coupling and the strong coupling expansions of the free energy are captured by the approximation, for all values of $\zeta$. The weak coupling expansion is an expansion in powers of $\la$. The one-loop result $F(\la=0)=0$ is reproduced by construction. From \eqref{Fpertsol} and \eqref{smalllabb}, we see that the two-loop result can be reproduced by choosing $\zeta$ such that
\be\label{twoloopsexact} \frac{8+\zeta(\zeta+2)^{2}(\zeta+8)}{4(\zeta+1)^{2}(\zeta+2)^{2}}=\frac{1}{2}\,\cvp\ee
which is
\be\label{twolexvalues} \zeta=\zeta_{1} = 0\quad\text{or}\quad \zeta=\zeta_{2}\simeq 3.5\, .\ee
The leading term in the strong coupling expansion of the bare bubble free energy always matches with the exact result \eqref{Fstrongsol}. The strong coupling expansion parameter is $1/\sqrt{\la}$ both in the approximate and exact result.

\noindent (ii) The bare bubble free energy does depend on the gauge-fixing parameter $\zeta$, but this dependence is under control, even for very large values of $\zeta$ for which it is given by \eqref{tildeFinf}. Preferred values of $\zeta$ correspond to the cases \eqref{twolexvalues} where the approximate and exact results match at two loops.

\noindent (iii) At the quantitative level, we find a remarkably precise match between the approximate and the exact results, all the way from the weakly coupled $\la\rightarrow 0$ to the very strongly coupled $\la\rightarrow\infty$ regimes, for a wide range of gauge-fixing parameters $\zeta$. For example, excellent results are obtained for the values \eqref{twolexvalues}, with a numerical error between the approximate and exact free energies never exceeding $4\%$ uniformly from $\la=0$ to $\la=\infty$! The boundaries in the space of all possible solutions, corresponding to the worst approximations above or below the exact result are found by solving the ``local'' gauge-independence condition
\be\label{zetagi} \frac{\partial\tilde F}{\partial\zeta} = 0\, .\ee
This yields either $\zeta=\zeta_{3}=\infty$ or some computable dependence $\zeta=\zeta_{4}(\la)$ between $\zeta$ and $\la$. These ``worst'' cases are actually found to be still quite reasonable.

\begin{figure}
\centerline{\includegraphics[width=6in]{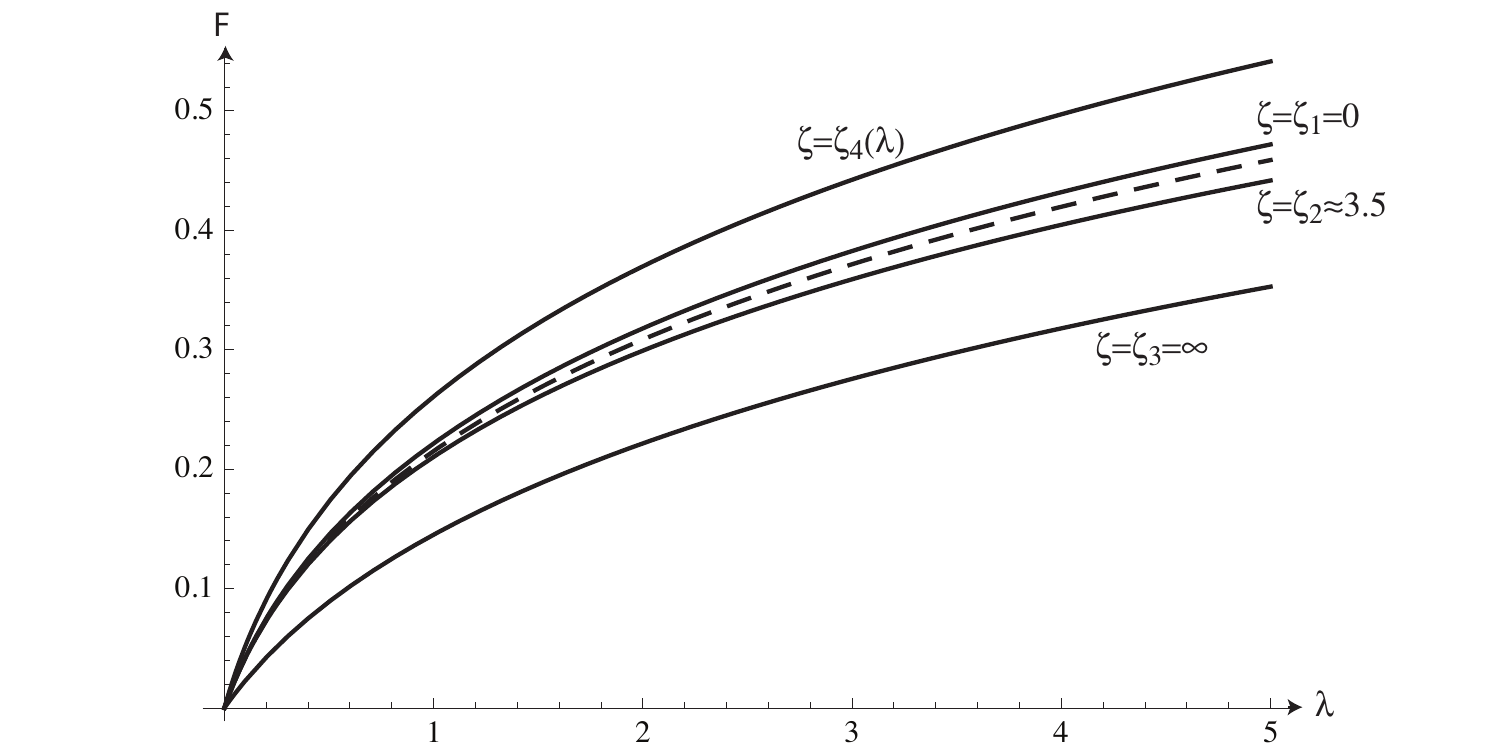}}
\caption{\label{fig10} Exact (dashed line) and bare bubble (plain lines) free energies as a function of the 't~Hooft coupling. The values $\zeta=\zeta_{1}=0$ and $\zeta=\zeta_{2}\simeq 3.5$ correspond to an exact match at two loops between exact and approximate solutions. The values $\zeta=\zeta_{3}=\infty$ and $\zeta=\zeta_{4}(\la)$ correspond to the boundaries of the region filled by drawing all possible approximate solutions obtained by varying $\zeta$ from zero to infinity. These boundaries  are also associated with the ``local'' gauge-independence condition \eqref{zetagi}.}
\end{figure}

\begin{figure}
\centerline{\includegraphics[width=6in]{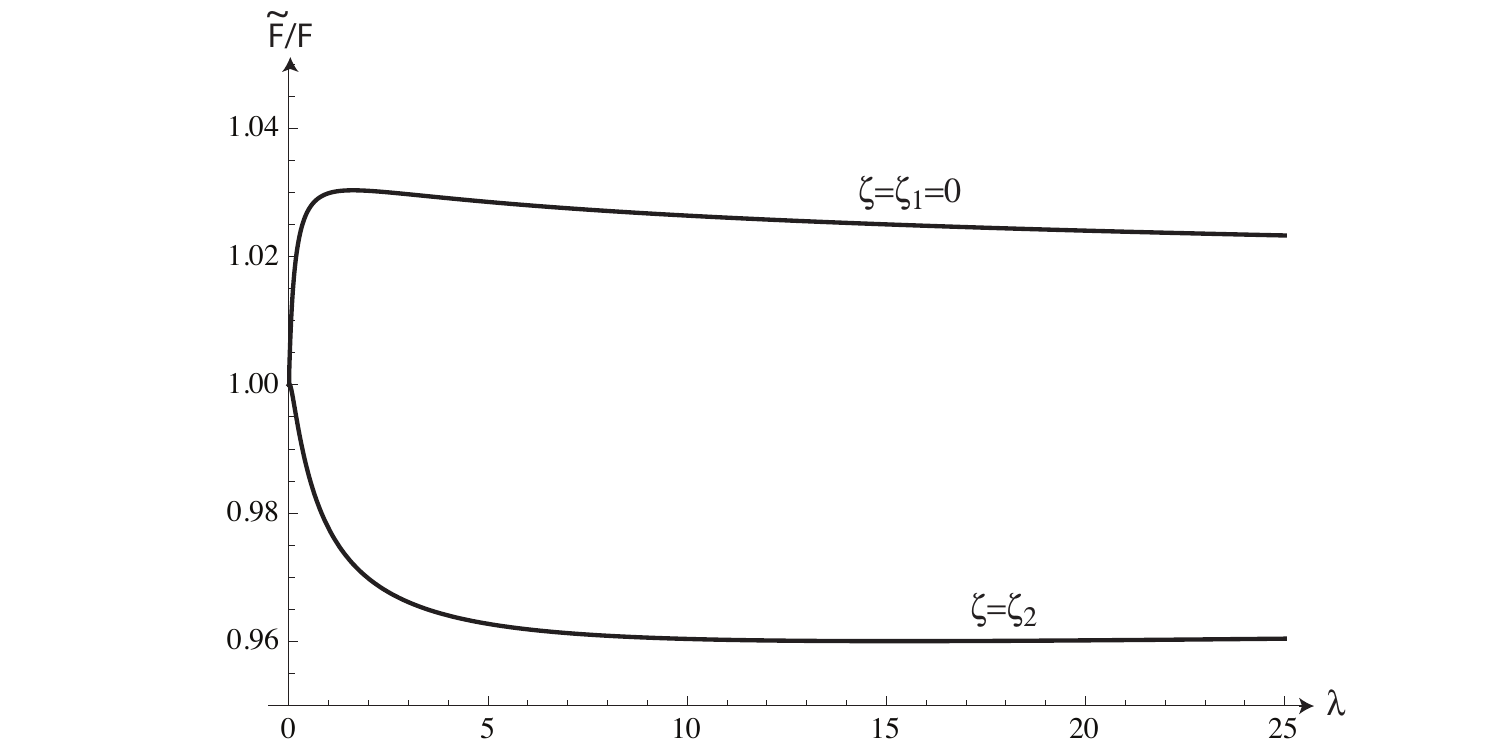}}
\caption{\label{fig11} Ratios between the bare bubble free energies, computed for the preferred values $\zeta=\zeta_{1}=0$ and $\zeta=\zeta_{2}$ of the gauge-fixing parameter, and the exact result. By definition of $\zeta_{1}$ and $\zeta_{2}$, the ratio goes to one when $\la\rightarrow 0$. From \eqref{Fstrongsol} and \eqref{stronglabb}, we know that the ratio also goes to one when $\la\rightarrow\infty$. Overall, the bare bubble approximation yields a remarkably accurate approximation to the exact free energy, with an error bounded by about $4\%$, uniformly in the 't~Hooft coupling.}
\end{figure}

\begin{figure}
\centerline{\includegraphics[width=6in]{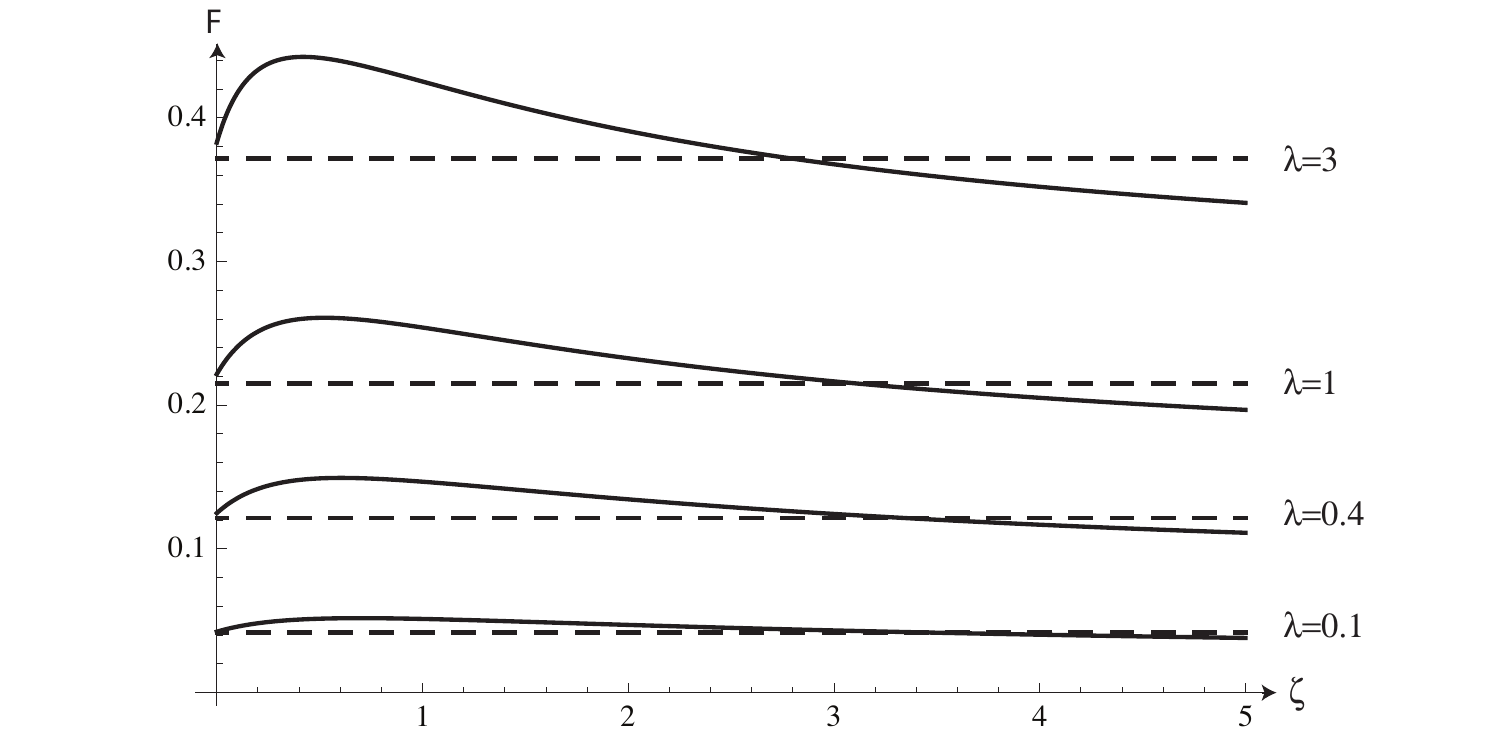}}
\caption{\label{fig12} Exact free energy (dashed lines) and bare bubble free energies as a function of the gauge-fixing parameter $\zeta$ (plain lines), for various values of the 't~Hooft coupling. The dependence in the gauge parameter $\zeta$ is rather mild and the approximation is excellent for a wide range of $\zeta$.}
\end{figure}

\noindent All these properties are illustrated on Figs.\ \ref{fig10}, \ref{fig11} and \ref{fig12}.

\subsection{The bare bubble correlators}

Let us now briefly discuss the correlators $C_{2k}=\langle\frac{1}{N}\tr M^{2k}\rangle$. The exact formula is given in \eqref{corrsol}. In the bare bubble approximation, we shall denote the correlators by $\tilde C_{2k}$. For simplicity, we limit ourselves to the gauge $\zeta=0$.

The correlators $\tilde C_{2}$ and $\tilde C_{4}$ can be obtained from \eqref{FtoCorr}, by using the formula \eqref{tildeFzero} for the approximate free energy,
\be\label{tildeC24} \tilde C_{2} = \frac{1}{4}\bigl(-1+\sqrt{1+8\la}\bigr)\, ,\quad \tilde C_{4} = \frac{1}{4}+\la-\frac{1}{4}\sqrt{1+8\la}\, .\ee
These formulas can be easily generalized to any correlator as follows. We add a source term $\frac{N}{\la}t_{k}\tr_{N} M^{2k}$ to the defining action $S_{N}$ of the model, such that
\be\label{C2ktk} C_{2k} = \la\frac{\partial F}{\partial t_{k}}\,\cdotp\ee
In the gauge $\zeta=0$, the resulting modification of the bare bubble probe brane action is simply to add the term $\frac{t_{k}}{\la} v^{2k}$. From \eqref{C2ktk} and \eqref{onshellA}, we thus get
\be\label{C2kbb1} \tilde C_{2k} = \frac{1}{2} v^{*2k}=2^{k-1}\tilde C_{2}^{k}= \frac{1}{2^{k+1}}\bigl(-1+\sqrt{1+8\la}\bigr)^{k}\, .\ee
This simple formula approximates very well the exact result \eqref{corrsol} for low values of $k$. For $1\leq k\leq 3$, the typical error is less than $10\%$, for all values of the 't~Hooft coupling. For larger values of $k$, and even though the error remains bounded for all values of $\la$, the approximation becomes less and less reliable. 

\section{Conclusion}

The one matrix model has always been extremely useful to clarify and illustrate some of the non-trivial aspects of the large $N$ limit in gauge theories. Following this philosophy, we have used it in this paper to exemplify the new approach to the large $N$ limit proposed in \cite{ferfund}, based on the study of D-brane probe actions. We have been able to compute the probe actions exactly in a family of equivariant gauges. We have checked explicitly several important aspects of the formalism, in particular the gauge invariance of the on-shell actions and their relation to the planar free energy. We have also shown that the gauge invariance was closely related to the spontaneous breaking of the equivariant BRST symmetry and to ghost condensation, phenomena which are likely to be quite generic and could play a crucial r\^ole in realistic models as well.

The above results clearly demonstrate the consistency of the approach proposed in \cite{ferfund}. It will be extremely interesting to apply the method to study more interesting gauge theories. Natural models to explore are supersymmetric gauge theories, for which many powerful techniques are available to derive exact results (see e.g.\ \cite{exacttech} and references therein). In particular, since the method makes the holographic properties of the gauge theories manifest, we could gain new interesting insights in cases where the holographic description remains rather mysterious, like for the $\nn=2$, $N_{\text f}=2$ superconformal Yang-Mills theory \cite{rastelli}. 

One of the most intriguing and exciting aspect of the whole construction is the existence of the bare bubble approximation. To our knowledge, it is the first time an analytic non-perturbative approximation scheme, that does not rely on the existence of a small parameter, is proposed in gauge theory. We have discussed in details the properties of the approximation in the one matrix model and found that it can yield remarkably accurate results, at least for some important observables like the free energy and some correlators. The most amazing feature is that the approximation can be reliable uniformly in the 't~Hooft coupling, all the way from the perturbative regime to the very strongly coupled regime. Needless to say, the properties of this approximation need to be investigated in more realistic models.

\subsection*{Acknowledgments}

This work is supported in part by the Belgian FRFC (grant 2.4655.07) and IISN (grants 4.4511.06 and 4.4514.08).

\begin{appendix}

\section{Appendix}

In this appendix, we derive from first principles the fundamental identity \eqref{ZNZN1A}, by using the path integral version of the equivariant gauge-fixing procedure explained in Section 4.3 of Ref.\ \cite{ferequiv}. 

Let us start from the definition \eqref{ZNdef} of the partition function, but for $N+1$ colors instead of $N$. We introduce an auxiliary Hermitian matrix $\mathcal H$ of size $(N+1)\times (N+1)$, in order to rewrite \eqref{ZNdef} in the equivalent form
\be\label{ZN1App} Z_{N+1}= \bigl(\la^{2}\zeta\bigr)^{-(N+1)^{2}/2}\int\!\d M\d H\, e^{-S_{N+1}(M)-\frac{N+1}{2\la\zeta}\tr_{N+1}\mathcal H^{2}}\, .\ee
The additional factor of $(\la\zeta)^{-(N+1)^{2}/2}$ in front of the integral is inserted to compensate the result of the Gaussian integral over $\mathcal H$. The right-hand side of \eqref{ZN1App} is thus manifestly independent of $\zeta$. (Note that we do not take care of trivial purely numerical $\la$- and $\zeta$-independent factors, which can be most easily fixed at the end of the calculation by looking at the classical limit.)

We then decompose the matrix $M$ as in \eqref{funddec} and use a similar decomposition for the matrix $\mathcal H$,
\be\label{Hdec} \mathcal H = 
\begin{pmatrix} H^{a}_{\ b} & \bar H^{a}\\ H_{a} & h
\end{pmatrix}\, .\ee
We then fix partially the gauge from $\text{U}(N+1)$ down to $\uN\times\text{U}(1)$ by imposing the Laudau-like conditions
\be\label{LanApp} w_{a}=H_{a}\, ,\quad \bar w^{a} = \bar H^{a}\, .\ee
As explained in \cite{ferequiv} (see also \cite{weinberg}), for partial gauge-fixing conditions imposed sharply in the path integral via a delta-function, the usual Faddeev-Popov trick applies straightforwardly.\footnote{The problem with more general $\xi$-gauge is that such gauges are obtained in the usual Faddeev-Popov procedure by introducing external classical fields. Such classical fields do not transform under the original gauge symmetry and thus always break the gauge group completely. Using classical external fields in the partial gauge-fixing procedure is thus inconsistent, since part of the gauge group must remain untouched. This difficulty is circumvented by introducing the auxiliary but dynamical field $\mathcal H$, on which the original gauge symmetry acts.} Introducing the ghosts and antighosts $\eta_{a},\bar\eta^{a},\chi_{a},\bar\chi^{a}$ associated with the conditions \eqref{LanApp}, we thus get from \eqref{ZN1App}
\begin{multline}\label{ZApp1} Z_{N+1} = \bigl(\la^{2}\zeta\bigr)^{-(N+1)^{2}/2}\int\!
\d v\d V\d w\d\bar w\d H\d h\d\eta\d\bar\eta\d\chi\d\bar\chi\,\\
e^{-S_{N+1}(M)-\frac{N+1}{2\la\zeta}\tr_{N+1}\mathcal H^{2}
-i(N+1)[ \eta_{a}((v-h)\delta^{a}_{b}-V^{a}_{\ b}+H^{a}_{\ b} )\bar\chi^{b} +\chi_{a}((v-h)\delta^{a}_{b}-V^{a}_{\ b}+H^{a}_{\ b})\bar\eta^{b}]}\, ,
\end{multline}
where now, taking into account \eqref{LanApp},
\be\label{Hdec2}\mathcal H = 
\begin{pmatrix} H^{a}_{\ b} & \bar w^{a}\\ w_{a} & h
\end{pmatrix}\, .\ee
We then use
\be\label{trH} \tr_{N+1}\mathcal H^{2} = \tr_{N}H^{2}+2w_{a}\bar w^{a}+h^{2}\ee
in \eqref{ZApp1} and perform the resulting Gaussian integrals over $H$ and $h$. This yields
\begin{align} \label{ZApp2}\begin{split} Z_{N+1}&=\bigl(\la^{2}\zeta\bigr)^{-(N+1)^{2}/2}\bigl(\la\zeta\bigr)^{(N^{2}+1)/2}\int\!\d V\d v\d w\d\bar w \d\eta\d\bar\eta\d\chi\d\bar\chi\\& 
e^{-S_{N+1}(M)-\frac{N+1}{\zeta\la}w_{a}\bar w^{a}-i(N+1)[\eta_{a}(v\delta^{a}_{b}-V^{a}_{\ b})\bar\chi^{b} + \chi_{a}(v\delta^{a}_{b}-V^{a}_{\ b})\bar\eta^{b}] - (N+1)\la\zeta(\eta_{a}\bar\eta^{a}\bar\chi^{b}\chi_{b}+\eta_{a}\bar\chi^{a}\chi_{b}\bar\eta^{b})}\end{split}\\\label{ZApp3}
\begin{split}
& = \la^{-\frac{N^{2}+1}{2}-2N}\zeta^{-N}\int\!\d V\d v\d w\d\bar w \d\eta\d\bar\eta\d\chi\d\bar\chi\\&\hskip 6cm e^{-S_{N+1}(M)-S_{\text{gauge-fixing}}[\zeta](V,v,w,\bar w,\eta,\bar\eta,\chi,\bar\chi)}\, ,\end{split}
\end{align}
where we have used the definition of the gauge-fixing action \eqref{gfaction} and \eqref{zetadef}. Using \eqref{stot0}, this equation can be immediately recast in the form
\be\label{ZApp4} \frac{Z_{N+1}}{Z_{N}} = \int\!\d v\, e^{-\mathscr A[\zeta](v)}\ee
for
\be\label{ZApp5} e^{-\mathscr A[\zeta](v)} = \la^{-\frac{1}{2}-2N}\zeta^{-N}e^{-(N+1)S_{1}(v)}\Bigl\langle\int\! \d w\d\bar w \d\eta\d\bar\eta\d\chi\d\bar\chi\, e^{-S_{N,1}[\zeta]}\Bigr\rangle\, .
\ee
The action $S_{N,1}[\zeta]$ is defined in \eqref{SN1form}, with $\xi=\la\zeta$, and the expectation value is taken over the matrices $V$ with the matrix model action $S_{N}(V)$. Introducing the auxiliary fields $\mu,\gamma_{1},\gamma_{2},\gamma_{4},\gamma_{4}$, we can express the result in terms of the action $\hat S_{N,1}[\zeta]$ defined in \eqref{hatSN1},
\begin{multline}\label{ZApp6} e^{-\mathscr A[\zeta](v)} =\la^{-\frac{1}{2}-2N}\zeta^{-N}
\la^{-\frac{1}{2}}(\zeta\la)^{-2}e^{-(N+1)S_{1}(v)}\\
\Bigl\langle\int\! \d\mu\d\gamma_{1}\d\gamma_{2}\d\gamma_{3}\d\gamma_{4}\d w\d\bar w \d\eta\d\bar\eta\d\chi\d\bar\chi\, e^{-\hat S_{N,1}[\zeta]}\Bigr\rangle\, .
\end{multline}
The additional factor $\la^{-1/2} (\zeta\la)^{-2}$ comes from the Gaussian integration over the auxiliary variables. Using the definition \eqref{AN1def} of the probe brane action, \eqref{ZApp6} is equivalent to
\be\label{ZApp7} e^{-\mathscr A[\zeta](v)} = \int\! \d\mu\d\gamma_{1}\d\gamma_{2}\d\gamma_{3}\d\gamma_{4}\,  e^{-\mathscr A_{N,1}[\zeta]}\, .\ee
The relation \eqref{ZNZN1A} that we wanted to prove then follows immediately from \eqref{ZApp4}. Note finally that, in the large $N$ limit, $\mathscr A_{N,1}[\zeta]\simeq N A[\zeta]$ and \eqref{ZApp7} yields
\be\label{Zapp8} \mathscr A[\zeta](v) \underset{N\rightarrow\infty}{=}
NA[\zeta](v,\mu^{*},\gamma_{1}^{*},\gamma_{2}^{*},\gamma_{3}^{*},\gamma_{4}^{*})\, ,\ee
where the star-spangled, on-shell values of the parameters are obtained by solving
\be\label{ZApp9} \frac{\partial A[\zeta]}{\partial\mu} = \frac{\partial A[\zeta]}{\partial\gamma_{1}} =\frac{\partial A[\zeta]}{\partial\gamma_{2}} =\frac{\partial A[\zeta]}{\partial\gamma_{3}} =\frac{\partial A[\zeta]}{\partial\gamma_{4}} =0\, .\ee

\end{appendix}
\end{document}